\documentclass{aa}  

\usepackage{lineno}
\usepackage{graphicx}
\usepackage{txfonts}
 \def\msun{\textrm{M}_{\odot}}
 \def\rsun{\textrm{R}_{\odot}}
\begin{document}

   \title{Asteroseismology of $\beta$ Cephei stars: The stellar inferences tested in hare and 
hound exercises}


   \author{S.J.A.J. Salmon
          \inst{1}
          \and
           P. Eggenberger \inst{1}
          \and
          J. Montalb{\'a}n\inst{2}
          \and
          A. Miglio \inst{2,3}
          \and
           A. Noels \inst{4}
                    \and
          G. Buldgen \inst{1}
                    \and
          F. Moyano  \inst{1}
                    \and
          G. Meynet \inst{1}
          }

   \institute{ Observatoire de Gen\`eve, Universit\'e de Gen\`eve, Ch. Pegasi 51, 1290 Sauverny, Switzerland \\
              \email{sebastien.salmon@unige.ch}
         \and
              Dipartimento di Fisica e Astronomia, Universit\`a degli Studi di Bologna, Bologna, Italy               
         \and
              INAF – Astrophysics and Space Science Observatory Bologna, Bologna, Italy               
         \and
              STAR Institute, Universit\'e de Li\`ege, All\'ee du 6 Ao\^ut 19C, 4000 Li\`ege, Belgium}
              
   \date{Received ;  }

 \titlerunning{Asteroseismology of $\beta$ Cephei stars}

 
  \abstract
   {The $\beta$ Cephei pulsators are massive, $\sim 8 - 25$M$_{\odot}$ essentially on the main-sequence, stars. 
The number of detected 
modes in $\beta$ Cephei stars often remains limited to less than a dozen of low radial-order modes. Such oscillation 
modes 
are in principle able to constrain the internal processes acting in the star. They probe the
chemical gradient at the edge of the convective core, in
particular its location and extension. They hence give constraints on macroscopic processes, such as hydrodynamic 
or magnetic instabilities, that have 
an impact on the mixing there. Yet, it is not clear to what extent the seismic inferences depend on the physics 
employed for the stellar modelling or on the observational dataset used. Consequently, it is not easy to 
estimate the accuracy and precision on the parameters 
and the nature of the physical processes inferred.}
   {We investigate the observational constraints, in particular the properties of the minimum set of pulsations 
detected, 
which are necessary to provide accurate constraints on the mixing processes in $\beta$ Cephei stars. We explore the 
importance of 
the identification of the angular degree of the modes. In addition, depending on the quality of 
the seismic dataset and the classical non-seismic constraints, we aim to estimate, in a systematic way, the 
precision achievable with asteroseismology on the determination of their stellar parameters.}
   {We propose a method extending the forward approach classically used to model $\beta$ Cephei stars. With the help of 
Monte-Carlo simulation, the probability distributions of the asteroseismic-derived stellar parameters were obtained. 
With these distributions, we provide a systemic way to estimate the errors derived 
from the modelling.  A particular effort was made to include, not only the observational errors, but also the 
theoretical 
uncertainties of the 
models. We then estimated the accuracy and precision of asteroseismology for $\beta$ Cephei stars in a series of hare 
and 
hound exercises.}
   {The results of the hare and hounds show that a set of four to five oscillation frequencies with an identified 
angular degree already leads to accurate inferences on the stellar parameters. Without the identification of the modes, 
the addition of other observational constraints, such as the effective temperature and surface gravity, still 
ensures the success of the seismic modelling. When the internal microscopic physics of the star 
and stellar models used for the modelling differ, the constraints derived on the internal structure remain valid if 
expressed in terms of acoustic variables, such as the radius. However, they are then hardly informative on structural 
variables expressed in mass. The characterisation of the mixing processes at the boundary of the 
convective core are model-dependent and it requires the use of models implemented with processes of a similar nature.}
   {}

   \maketitle
%

\section{Introduction}

The $\beta$ Cephei stars are pulsating stars of masses between $\sim$8 and 
25~M$_{\odot}$. 
Their pulsations are low-order pressure (p) and gravity (g) modes with periods typically of $\sim$0.5 to 8~hours. Since 
part of these $\beta$ Cephei modes present a mixed p- and g- character, they are privileged 
targets to test physical processes at the boundary of the convective core and radiative envelope with asteroseismology. 
A series of 
asteroseismic modellings have succeeded in determining their stellar parameters and core overshoot in a dozen of 
$\beta$ 
Cephei stars \citep[see reviews by][]{aerts13,aerts15,bowman20}, and in at least four of them, the core and surface 
rotation 
rates \citep[see review by][]{goupil11}.

The question of mixing of chemical species at the border of convective cores, for example by core overshoot, is of 
prime 
importance for stellar evolution \citep[e.g.][]{maeder76,pinsonneault97}. The isochrone fitting of stellar clusters 
\citep[see][for 
a review]{gallart05}, calibration of eclipsing binaries \citep[e.g.][]{ribas00,claret17}, or low-mass star 
asteroseismology \citep[e.g.][]{Miglio07,deheuvels16} have indeed revealed the need for such extra mixing. The 
extra mixing near the convective core can significantly affect the age determination of 
stellar clusters, leading to uncertainties between $\sim$40 up to 300\%, depending on the mass of stars at 
the turn-off \citep{meynet09}. In stars with masses $M\gtrsim$7-8 M$_{\odot}$, 
quantifying the extra mixing during the main sequence (herafter MS) is also essential to determine their subsequent 
phases of evolution and 
their contribution to nucleosynthesis \citep[e.g.][]{chiosi86}. Many overshooting prescriptions 
exist \citep[see reviews by][]{chiosi06,salaris17}, with the latest attempts being based on the results from
three-dimensional (3D) simulations, as in \citet{scott21}. Despite these efforts, there is still no clear 
evidence on which 
one to adopt. In addition to overshooting, mixing induced by rotation is also expected to contribute to core 
extra mixing \citep[see][in the case of early-type stars]{talon97}. Internal gravity waves generated by convective 
motions at the interface of the core might also induce the mixing of chemical species 
\citep[e.g.][]{press81,montalban94,talon08,rogers17}. 
The asteroseismology of MS B stars remains a crucial testbed for confronting these physical 
prescriptions. 

Recently, the detection of gravity mode period-spacing in SPB stars, which are pulsating late B-type stars of 
intermediate masses, has delivered information on the nature and extent of this extra mixing 
\citep[e.g.][]{degroote,iran16,michielsen21,may21}. This has confirmed the potential of the asteroseismology of 
SPBs as 
anticipated in theoretical 
studies \citep{miglioms,miglioms2,may18,aerts18}. Similarly, the $\beta$ 
Cephei stars can in principle unveil constraints on the nature of this mixing in more massive 
stars \citep[][]{migliocom,montalban08,michielsen19}. However, the degree to which it depends on the 
stellar models used in the asteroseismic modelling needs more investigation. Asteroseismic constraints on this 
extra mixing have been retrieved as an overshooting parameter ($\alpha_{\textrm{ov}}$) in a dozen of $\beta$ Cephei 
stars \citep[see reviews by][]{aerts06,aerts15}. The value of masses and 
overshooting determined for different $\beta$ Cephei with asteroseismology show a large dispersion in the values of the 
overshooting\footnote{stars at different evolutionary stages on the main sequence may explain part of the 
dispersion} and errors on the masses fluctuating from a few to 40 \%. The overshooting values are dependent 
of the formalism used in each study \citep[see also][]{martinet21}, since they correspond to the overshooting parameter 
of the stellar 
models that best fit the asteroseismic observables. The errors will vary depending of the method (including stellar 
models) and set of observational constraints, and hence without a systematic 
approach, it remains difficult to estimate the real accuracy and precision one can expect in the modelling of 
$\beta$ Cephei stars. Moreover,  \citet{dziembowski4} tried in 
a more complex attempt to determine the shape of the chemical composition gradient in the overshoot region for 
the $\nu$ Eri and 12 Lac stars but they could not draw clear conclusions.

With the large collection of data for B-type pulsators obtained from the ground, either in dedicated or large surveys  
\citep[see e.g.][]{handler,handler11,ngc6910,labadie20} and the results from space missions CoRoT, \emph{Kepler} and
TESS \citep{degrootecorot,balonakepler,may19b,burssens20,sze21}, 
the number of known $\beta$ Cephei pulsators has grown rapidly in the last years. With the dedicated BRITE space 
mission, we also benefit of rich datasets for the analysis of $\beta$ Cephei stars \citep[][]{handler17,walzak19}, 
leaving room for new progresses. The actual potential of asteroseismology, and the precision and accuracy it offers 
on the inferred stellar parameters needs to be 
addressed. The information on the stellar 
parameters and internal structure depends crucially on the nature of the observations: on the one hand, the seismic 
data, which includes the number, precision, and identification of detected modes, and on the other hand, the classical 
observables such as the effective 
temperature, surface gravity, and photospheric chemical abundances. The seismic inferences also 
depend on the physics adopted in the theoretical models: micro-physics -- the choice of the opacity and chemical 
mixture--, and macro-physics describing the extra mixing at the core boundary. We propose here to explore how and 
which of these factors are 
critical to the success of the asteroseismic modelling of $\beta$ Cephei targets (see also preliminary studies on 
$\beta$ Cephei stars by \citealt{thoul03-2} or sdB pulsators by \citealt{vvg08}).  

To answer these questions, we carried out a series of hare and hound exercises, so extending the effort 
initiated 
in \citet{migliohh}. The importance of working with a well characterised seismic dataset has also been explored very 
recently in \citet{bowman21} for SPB pulsators. We proceeded by computing theoretical stellar models that then 
served as simulated observed stars on which an 
asteroseismic 
modelling was then performed. We have introduced a method based on Monte-Carlo simulations to assess the reliability of 
the solution: for a given observational set of frequencies, we generated new sets of frequencies drawn from 
Gaussian distributions centred on the observed ones. The width of the distributions accounts for theoretical 
uncertainties and observational errors. For each set generated, a new best-fit 
model was determined following an independent asteroseismic modelling. Gathering the results for each generated set of 
frequencies, we can construct 
a distributions of the inferred stellar parameters and use it to estimate reliable errors of the modelling process. 
With the help of this re-sampling method, we explored how the inferences rely on the seismic and classical constraints 
by varying the properties of the models serving as simulated stars and the set of observational constraints. We then 
looked at the effect of various physical assumptions:  the importance of the chemical mixture adopted in the 
models and the sensitivity to the nature of extra mixing processes.

The paper is divided as follows: Sects.~\ref{section2} and ~\ref{section3} describe the asteroseismic method we 
have 
developed and the input physics 
of the grid of models used for the modelling, respectively. Section~\ref{section4} presents the 
simulated targets and the goals of the different hare and hound exercises. The results of the exercises are 
presented and commented in Sect.~\ref{section5}, assessing the role of the seismic data and of the 
physics assumed in the theoretical models. We then synthesise the main results of the paper in the conclusion.

\section{The asteroseismic method}
\label{section2}

The asteroseismic modelling of $\beta$ Cephei stars is commonly based on a forward approach by a direct comparison of 
the 
observed frequencies to those of 
theoretical stellar models. Our method is first based on the same approach, using the following seismic 
merit function: 
\begin{equation}
 \chi^2=\frac{1}{N_{\mathrm{obs}}}\sum_{i=1}^{N_{\mathrm{obs}}} 
\frac{(\nu_{\mathrm{obs,i}}-\nu_{\mathrm{th,i}})^2}{\sigma_{i}^2},
\label{eq1-section2}
\end{equation}
where $N_{\mathrm{obs}}$ is the number of observed frequencies, $\nu_{\mathrm{th,i}}$, an adiabatic theoretical 
frequency, $\nu_{\mathrm{obs,i}}$ and $\sigma_{i}$ are an observed frequency and its associated uncertainty, 
respectively. 
Foreseeing the large set of fundamental parameters typical of $\beta$ Cephei stars, given they are expected 
to span a large mass 
interval, around 8 up to 25 M$_{\odot}$ \citep[see the case of HD 46202][]{briquet11}, we make use of a grid 
of pre-computed 
stellar models (see details in Sect.~\ref{section3}).  For each model, the set of $\nu_{\mathrm{th,i}}$ that 
minimise the distance $|\nu_{\mathrm{obs,i}}-\nu_{\mathrm{th,i}}|$ is chosen to compute Eq.~(\ref{eq1-section2}). 

When a mode identification is available, either its angular degree $\ell$ and/or its azimuthal order, $m$, 
we require the $\nu_{\mathrm{th,i}}$ that matches $\nu_{\mathrm{obs,i}}$ to be of the same $\ell$ and/or $m$. If two or 
more $\nu_{\mathrm{obs,i}}$ are 
associated with a same $\nu_{\mathrm{th,i}}$, we discard the stellar model. The $\chi^2$ function for all of the 
stellar models 
composing 
the grid is then computed, and the best-fit model is taken as the one associated with the global minimum in $\chi^2$.

Our theoretical dataset of oscillation frequencies is restricted to modes with angular degree $\ell \leq 3$. 
Theoretical 
computations show that the most visible modes (from photometric or spectroscopic 
detections) are limited to degrees up to $\ell=$3--4 \citep[see e.g. Figs. 6.4 and 6.14 in][]{aertsetal}. Despite the 
fact that the $\ell$=3 modes are those suffering the most of geometric cancellation effects in photometric data, they 
remain 
detectable in spectroscopic observations \citep[see the case of V2052 Oph in][]{briquet12}. We note that in 
some cases the 
situation could be different: for instance, CoRoT results 
revealed modes with identifications claimed up to $\ell$=9 in intermediate-mass stars such as $\delta$ Scuti 
pulsators \citep{mantegazza12}.

We 
do not include 
the effective temperature ($T_{\mathrm{e}}$) and surface gravity ($\log g$) as constraints in the merit function. If 
they 
are to be used as additional constraints, instead we select the best-fit model as the local minimum falling in either 
the 1- or 3-$\sigma$  error box on these two observables. Hereafter we refer to 1-$\sigma$ or 3-$\sigma$ 
constraints to mention that we require the solution to be part of the 1- and 3-$\sigma$ error boxes, 
respectively. The metallicity ($Z$) can also be used as an additional classical constraint, similarly to 
$T_{\mathrm{e}}$ and $\log g$. Yet, we remarked that in most of the cases we 
have treated hereafter, it did not improve the results (adopting an error on $Z$ representative of the 
typical observational errors for hot stars).

\subsection{Evaluating the precision of the fit}
\label{section2-1}
As defined in Eq.(\ref{eq1-section2}), the $\chi^2$ function would follow a $\chi^2_{\mu}$ statistic with $\mu$ 
degrees of freedom, provided it respects two conditions: first, the frequencies $\nu_{\mathrm{obs,i}}$ have to be 
independent and secondly, the theoretical model has to be expressed by a linear 
combination\footnote{or at least does not 
depart too strongly from linearity} function of the parameters, $a_j$, which are adjusted to obtain the best fit to the 
observations.  These conditions are not met in the case of $\beta$ Cephei pulsators. Indeed, the 
frequencies $\nu_{\mathrm{th,i}}\equiv\nu_{\mathrm{th,i}}(a_1,...,a_j,...a_n)$, 
rely on global quantities symbolised by the $n$ parameters $a_j$, with $n=5$. These five parameters are $M$, $X$, 
$Z$, $\alpha_{\mathrm{ov}}$, and the age, with $X$ the initial hydrogen mass fraction, and $Z$ the initial metallicity. 
For instance, in the simple case of 
radial oscillation modes, the relation of their frequencies $\nu_{\mathrm{th}} 
\propto 1/\tau_{\mathrm{dyn}}$, with the dynamical timescale $\tau_{\mathrm{dyn}}=\sqrt{R^3/GM}$ ($R$ being the radius, 
and $G$ the gravitational constant), already reveals a non-linear 
dependence on $M$. 

\begin{figure}[!]
\centering
\includegraphics[width=0.5\textwidth]{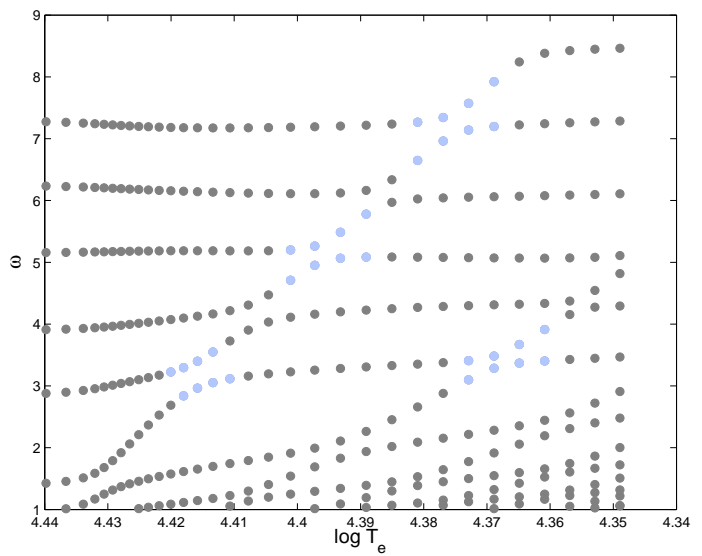}
\caption{Evolution along the main sequence of the dimensionless frequencies of $\ell$=1, $k<5$ modes in a 
11~M$_{\odot}$ model. Four avoided crossings occurring during this sequence are highlighted in light blue. }
\label{fig-avoided}
\end{figure}

More generally, due to the mixed character of the low radial order modes (using $k$ for the radial order hereafter) in 
$\beta$ 
Cephei stars, the frequency spectra present features known as \emph{avoided crossings}. The mixed modes present both an 
acoustic (p-) 
and gravity (g-) mode behaviour and appear when the 
propagation cavities of p- and g-modes are only separated by a small evanescent region. As the $\beta$ Cephei stars 
develop a 
chemical composition gradient as their convective core recedes, a sharp and localised peak in the Brunt-V\"ais\"al\"a 
frequency appears, favouring the presence 
of mixed modes.  The avoided-crossing phenomenon is shown in Fig.~\ref{fig-avoided} in light blue in the 
main-sequence evolution of the dimensionless \footnote{$\tau_{\textrm{dyn}}$ is used for the 
nondimensionalisation} frequencies ($\omega$) of low-radial order modes of a typical $\beta$ 
Cephei 
model. 
The sharp variations of the frequencies induce a non-linear relation between the frequencies and the parameters of the 
model \citep[e.g.][]{scuflairemixed,shiba79}.
 
Thus we use a re-sampling method to 
explore the accuracy and precision on the stellar parameters derived from the selected best-fit model. We perform 
Monte-Carlo simulations by randomly drawing frequencies that we call pseudo-observed frequencies. Assuming 
$N_{obs}$ frequencies have been observed for a given star, we build $M_{\mathrm{sim}}$ new sets, each composed of 
$N_{obs}$  pseudo-observed frequencies. These latter are drawn one-by-one from Gaussian distributions of standard 
deviation, $\sigma_{\mathrm{\nu,i}}$ centred on the $\nu_{\mathrm{obs,i}}$. The values of $\sigma_{\mathrm{\nu,i}}$ are 
tailored not only to account for the observational errors, but also the uncertainties in the 
physics of the models. A natural choice is hence to set $\sigma_{\mathrm{\nu,i}}=\max 
(\sigma_{\nu_{\mathrm{obs,i}}},\sigma_{\nu_{\mathrm{theor,i}}})$, where $\sigma_{\nu_{\mathrm{obs,i}}}$ and 
$\sigma_{\nu_{\mathrm{theor,i}}}$ are the observational and theoretical errors, respectively (see 
details in Sect.~\ref{section2-3}). 

A best-fit stellar model (based on the minimum in $\chi^2$) is then derived for each of the $M_{\mathrm{sim}}$ 
simulated sets of pseudo-observed frequencies. The stellar parameters of the best-fit model 
 of each simulation are 
collected. The distributions of stellar parameters obtained in this way are then considered as the distributions of the 
solution in the space of parameters, accounting for the observational errors and 
theoretical uncertainties of the models (see Sect.~\ref{section2-3}). 

Indeed, we read these distributions built from the simulations as 
probability density distributions. The medians of the 
marginal distribution of a given stellar parameter are picked up as the inferred results from the Monte-Carlo 
(MC hereafter) simulations. We also derive confidence intervals on the estimated stellar
parameters. The lower and 
upper limits of the confidence intervals at the 1-$\sigma$ level, conventionally set to 68.3\%, correspond to the 
$(1-\sigma)/2\times M_{\mathrm{sim}}$ and $(1+\sigma)/2\times M_{\mathrm{sim}}$ values of the ordered marginal 
distribution. 

We can then compare these confidence intervals to the single set of stellar parameters obtained from the minimum in 
$\chi^2_{\mathrm{min,true}}$, which in this case is based on the actual observed frequencies.

By carrying out tests with assumed different numbers of 
observed frequencies (keeping it representative of typical $\beta$ Cephei observations)  and numbers of simulations, we 
found that increasing the number of simulations above $M_{\mathrm{sim}}$=1000 did not change the estimated errors nor 
did they induce changes in the distributions of the inferred stellar parameters. We consequently adopted that value in 
our 
different modelling exercises.


\subsection{Parameters of the Monte-Carlo simulations}
\label{section2-3}

To determine the value of $\sigma_{\mathrm{\nu,i}}$ to run the simulations, we review the expected observational and 
theoretical errors affecting $\beta$ Cephei frequencies. The observational errors are typically of 10$^{-4}$-10$^{-6}$ 
cycle per day (c/d) for well-studied $\beta$ Cephei stars, see for instance \citet{nuerispec} and \citet{nueriphot} in 
the case of ground-based spectroscopy and photometry, respectively; or \citet{spacephot} for the case of 
space-based photometry. 

Different theoretical uncertainties contribute to the theoretical error. We first tested the uncertainty in the 
micro-physics of the model by comparing the adiabatic frequencies between stellar models with identical stellar 
parameters, but differing in the chemical mixture and opacity. Their theoretical pulsation frequencies varied 
by $\sim 10^{-3}$--10$^{-2}$~c/d. On the scale of macro-physics, stellar rotation, not included in the models of 
our grid, is likely one of the processes potentially impacting at most the stellar structure, but also impacting 
directly the physics of oscillations. Including the effect of rotation on oscillations and on the stellar 
structure (i.e. distortion), \citet{theorerr} found a theoretical difference up to 
$\sim$~5$\times$~10$^{-2}$~c/d.

The computed frequencies of oscillation also depend whether the energy equation is 
treated under the adiabatic assumption. However, given the non-adiabaticity of superficial 
stellar layers, a non-adiabatic approach can reveal shifts in the frequencies. We hence computed adiabatic and 
non-adiabatic frequencies for a sample of models representative of the theoretical grid. We used for this purpose the 
LOSC ( Li\`{e}ge OScillation Code) adiabatic and MAD non-adiabatic codes, in their 
default setting mode, as detailed respectively in \citet{losc} and \citet{mad}.  We found differences between 
adiabatic and non-adiabatic frequencies from $\sim$~10$^{-4}$ to 10$^{-2}$ c/d, depending on the $\ell$ degree and 
evolutionary stage. We finally took the maximum of these differences for $\sigma_{\nu_{\mathrm{theor,i}}}$, that is 
10$^{-2}$~c/d, as the value representative of the theoretical uncertainties. The observational errors on frequencies 
are typically two to three orders smaller than the theoretical ones, so that the limits on asteroseismic precision 
shall be 
dominated by current model uncertainties.

\subsection{Typical observational errors on the fundamental parameters}
\label{section2-4}

To test the potential of classical constraints, we need to estimate what are the typical values of their 
observational errors in the case of $\beta$ Cephei stars. The fundamental parameters are usually derived from 
spectroscopic 
observations. In particular, the effective temperature of early B-type stars can be determined from the ionisation 
balance 
of spectroscopic lines of He, C, O, Ne, Si or Fe. When a maximum of these indicators is detected, the error on 
$T_{\mathrm{e}}$ can be as low as 200-300 K \citep[e.g.][]{nieva12}. However all these lines are not always 
available: instead strong lines of Ne and Si are then used, leading to an error of $\sim$1000 K (T. Morel, private 
communication). We adopt as the error on $T_{\mathrm{e}}$ an intermediate value of 700~K for the hare and hound 
exercises. The surface gravity $\log g$, is estimated with the help of a fit to the wings of H Balmer lines. In that 
case, a minimum 
0.15 dex reasonably accounts for the possible sources of error (T. Morel, private communication). We 
note that depending of the cases, for examples in reason of fast rotation or radial velocity variability in a binary 
system, the error can be larger, around 0.30 dex.

The spectra of early-type stars 
present many lines 
of N, O and Fe, and to a lower extent of C, enabling to determine their photospheric abundances. Those of Ne, Mg, Si, 
and 
S can also be detected under favourable conditions, then giving an insight on the most abundant elements. The 
present-day ratio 
$Z/X$ can then be calculated by assuming the abundances of minor elements follow the solar mixture. Different studies 
have revealed that abundances of B stars in the solar neighbourhood differ and are metal poorer than the solar 
abundances \citep{norbert,morel09,niem09,nieva11}. The difference was particularly marked with past solar 
abundance 
determinations such as the  \citet{gn93}, hereafter GN93, one. Revision of solar abundances led to an important 
decrease of the solar metallicity in 2005 \citep[][hereafter 
AGS05]{asplund}. A subsequent revision \citep[][]{AGS09} has moderated this downward revision of the solar 
metallicity, but the AGS05 appears today the more representative of the chemical mixture of neighbour B stars. Once 
the solar distribution assumed, obtaining the metallicity from $Z/X$ still requires to determine $X$, for instance from 
a solar 
calibrated model. \citet{morel06} derived the metallicity in this way for a 
sample of $\beta$ Cephei stars, finding an average error on $Z$ of $\sim$ 0.002.

If the number of individual abundances is insufficient, metallicity is derived from the relation [Fe/H]=[$Z/X$] that 
assumes the stellar abundances are distributed following the solar mixture. In the literature we find errors on the Fe 
abundance of early-type stars from $\sim$0.1 to 0.2 dex \citep[e.g.][]{morel06,nieva12}. This translates to 
typical errors on the metallicity $Z$ of $\sim$0.003 and 0.005. As we see in the next sections, the uncertainty on 
the 
metallicity derived with help of the MC simulations are typically of $\sim$0.002--0.004, except for of a few 
cases.

\section{Theoretical grid of models and their frequencies of oscillation}
\label{section3}

The stellar parameters covered by the grid of models are detailed in Table \ref{table1-section3}. The grid was designed 
at first for the modelling of 
the $\beta$ Cephei star HD 180642 by \cite{reftogrid} and was computed with the Li\`{e}ge stellar evolution 
code \citep[CLES,][]{cles}. The grid is not purposed for stars affected to a subtantial 
level by rotation and so the models do not include rotational effects. The treatment of convection follows 
the mixing-length prescription by \citet{cox68}, with its 
parameter solar calibrated at $\alpha_{MLT}$~=~1.8. Evolutionary tracks with or without overshooting are 
considered (see Table~\ref{table1-section3} for the values of the overshoot parameter). The mixing in the overshoot 
region is treated as instantaneous. The surface boundary conditions are obtained from Eddington's law ($T[\tau]$) for a 
grey atmosphere. The nuclear reaction rates are those of \citet{caughlan} with the revised $^{14}{\rm 
N}(p,\gamma)^{15}{\rm O}$ cross section from Formicola et al. (2004)\nocite{formicola}. The solar chemical mixture 
of AGS05 is adopted (see Sect.~\ref{section2-4}). Opacities corresponding to this chemical mixture are computed with 
the OP tables \citep{badnell}, while the updated OPAL equation of state from \cite{rogers} is used. The full 
set of input parameters covered by the grid, which are the stellar mass ($M$), the initial hydrogen mass fraction 
($X$), 
the initial metal mass fraction ($Z$), and the overshooting parameter ($\alpha_{\mathrm{ov}}$), are given in 
Table~\ref{table1-section3}. 

\begin{table}[!h]
\caption{Stellar parameters of the grid.}             
\label{table1-section3}      
\centering                          
\begin{tabular}{c c c }        
\hline\hline                 
Parameter & Range & Step  \\    
\hline                        
   M (M$_{\odot}$) & 7.6 -- 18.6 & 0.1 \\
   X & 0.68 -- 0.74 & 0.02  \\
   Z & 0.010 -- 0.018 & 0.002  \\
   $\alpha_{\mathrm{ov}}$ & 0 -- 0.50 & 0.05 \\      
\hline                                   
\end{tabular}
\end{table}

\begin{figure}[!]
\centering
\includegraphics[width=0.49\textwidth]{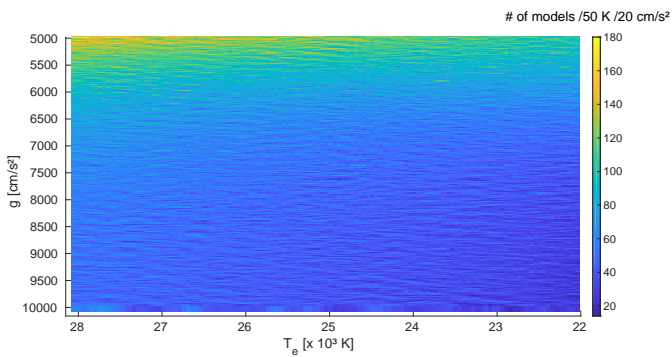}
\caption{Density of models constituting the theoretical grid. Each mesh of the figure covers 50~K in $T_{\textrm{e}}$ 
and 20~cm/s$^2$ in surface gravity. We scan the full range of input parameters as given in Table 
\ref{table1-section3}. The yellower a mesh block is, the higher is the number of models belonging to it, as reported 
in the legend on the right.}
\label{fig1-section3}
\end{figure}

\subsection{Oscillation frequencies}

The adiabatic frequencies of oscillation are determined for all the models from the Zero Age Main Sequence (ZAMS) to 
the 
Terminal Age Main Sequence (TAMS), with the help of the LOSC tool, adopting the standard 
boundary conditions of the code \citep[see exact details in][]{losc}. For each model 
on the main sequence, the radial ($\ell$=0) and non-radial ($\ell$=1,2,3) low-order p- and g-modes are computed. 
 For each model and each $\ell$, the modes are computed in the same fixed interval of adimensional 
frequencies, chosen as representative of $\beta$ Cephei stars.

\subsection{Resolution of the grid}
\label{section3-1}
The density of models in a restricted region of the grid is shown in Fig.~\ref{fig1-section3} (the whole grid ranges in 
$T_{\textrm{e}}$ from $\sim$~14,400 to $\sim$~36,000 K and surface gravities from 260 to 22,400 cm/s$^2$, or $\log g 
\sim$ 2.40 -- 4.35). There 
are typically several dozens of models per 50 K and per 20 cm/s$^2$ boxes. The denser parts correspond to models in the 
end of the main sequence, and which overlap with less massive models presenting higher values of 
$\alpha_{\textrm{ov}}$. Overshooting not only extends the MS lifetime but increases as well the luminosity of 
evolutionary tracks at the end of MS so that the models with high overshooting cross the tracks of more massive ones.

\section{Selection of the exercises}
\label{section4}

The selection of the target stellar models (i.e. the hares) analysed in the different exercises is presented in Table 
\ref{table1-section4}. The stellar parameters of the models were chosen to be fully representative of typical $\beta$ 
Cephei stars, in terms of mass and evolutionary stage. As a proxy of the age, we use the central hydrogen mass 
fraction, $X_{\textrm{c}}$, in the parameters describing our models and results. The goal of the modelling is to 
retrieve the information on the target stellar model and look at the quality of the derived inferences.
Hereafter, we 
use the term ``input parameters'' of the exercise to refer to the simulated set of observations (the hare): the results 
of the modelling are mentioned as the output or derived parameters. As requested in hare and hound 
exercises to avoid any bias in the modelling and interpretation of the results, the computation and selection of the 
input frequencies were done by a different person than the one in charge of the asteroseismic modelling and analysis of 
each exercise.

\begin{table*}[!]
\caption{Stellar parameters of the target modes used as hare in the hare and hound exercises.}             
\label{table1-section4}      
\begin{center}

\begin{tabular}{c c c c c c c c c c c c c}     
\hline\hline       
Model  & M & R & X & Z & $\alpha_{\mathrm{ov}}$ & $X_{\mathrm{c}}$ & Diff & Solar mixture & Opacities & Atm &  
$T_{\mathrm{e}}$ & $\log g$  \\ 
  & ($M_{\odot}$) & ($R_{\odot}$) & & & & & & & & & (K) & ($g$ in cm/s$^2$) \\ 
\hline
 t1 & 14 & 7.48 & 0.70 & 0.014 & 0.2 & 0.288 & N & AGS05 & OP & Edd & 27647 & 3.8364 \\                  

 t2 & 11 & 5.98 & 0.70 & 0.016 & 0.2 & 0.351 & N & GN93 & OP & K & 25293 & 3.9258 \\

t6 & 10 & 4.020 & 0.70 & 0.014 & 0 & 0.388 & Y & AGS05 & OP & Edd & 24487 & 4.0196 \\

\hline                  
\end{tabular}
\end{center} 
{\flushleft The acronyms GN93 and AGS05 stand for the solar chemical mixtures of \citet{gn93} and \citet{asplund}, 
respectively. The ``Diff'' column indicates whether the model 
is computed with additional turbulent mixing (Y) or without (N). The Edd and K symbols indicates whether the stellar 
model atmosphere is computed following Eddington's law or from Kurucz models.}
\end{table*}

The list of the different exercises and the input frequencies are summarised in Table \ref{table2-section4}. The names 
of the exercises begin with the name of the input model and end with a short indication 
on the input constraints and stellar physics that are explored in the exercise. 

\begin{table*}[!]
\caption{Frequencies of the target models used as hare for the hare and hound exercises.}            
\label{table2-section4}      
\begin{center}
\begin{tabular}{l l}     
\hline\hline       
Exercise label  & $\nu_{\mathrm{obs,i}}$ (c/d) \\ 
\hline                    
 t1-asp-3freq$^\dag$ & 5.195081 ($\ell=0$) ; 5.634114 ($\ell=1$) ; 7.807939 ($\ell=2$) \\ 
 t1-asp-4freq$^\dag$ & 5.195081 ($\ell=0$) ; 5.634114 -- 8.668213 ($\ell=1$) ; 7.807939 ($\ell=2$) \\
 t1-asp-5freq$^\dag$ & 5.195081 ($\ell=0$) ; 5.634114 -- 8.668213 ($\ell=1$) ; 7.807939 -- 8.290281 ($\ell=2$) \\
 t1-$\ell$1-wrong$^\dag$ & same as t1-asp-5freq but with $\nu=$8.668213 identified as $\ell=2$\\
 t1-$\ell$-4unknown$^\dag$ & same as t1-asp-4freq but without knowledge of $\ell$ \\
  t1-$\ell$-5unknown$^\dag$ & same as t1-asp-5freq but without knowledge of $\ell$ \\
 t2-gn93$^\dag$ & 6.461264 -- 8.364639 ($\ell=0$) ; 6.927369 -- 9.227121 ($\ell=1$) ; 8.768298 -- 10.152126 
($\ell=2$) 
\\
 t6-asp-diff & 7.733229 -- 9.972529 ($\ell=0$) ; 5.364213 -- 8.340911 -- 11.159961 ($\ell=1$) ;\\
 & 7.353429 -- 8.577249 -- 9.818674 ($\ell=2$) \\
\hline                  
\end{tabular}
\end{center}
{$\dag$ The frequencies of the target model are computed with the non-adiabatic 
oscillation code MAD.}
\end{table*}

The purpose of the first series of exercises (six first rows of Table~\ref{table2-section4}) is to explore 
the quality of the asteroseismic inferences, based on the seismic dataset at disposal: number of detected frequencies, 
known or 
unknown 
determination of the mode degrees, error in the degree identification of a mode \citep[see also similar work 
by][in the case of SPB stars]{bowman21}. As overshooting is 
included in the 
models of the grid, the degrees of freedom of the problem are the five stellar parameters: $M$, $X$, $Z$, 
$\alpha_{\mathrm{ov}}$, and the 
age (here the proxy $X_{\textrm{c}}$). However, in some cases, the number of axisymmetric mode frequencies detected in 
$\beta$ Cephei is lower, down to 3 as for examples in HD 129929 or V2052 Oph 
\citep[][respectively]{aerts03,briquet12}. So we 
started considering three as the assumed number of detected frequencies in the t1-asp-3freq series of exercises, and 
increased it to four and five frequencies in the t1-asp-4freq and t1-asp-5freq ones. Moreover, in the 
t1-$\ell$-4unknown and t1-$\ell$-5unknown exercises, we considered the same four and five modes as in the t1-asp-4freq 
and t1-asp-5freq cases, but assuming no knowledge 
of the mode angular degrees. In the t1-$\ell$1-wrong, the effect of a misidentification of the angular degree of one 
mode is considered, based on the same five frequencies than in the t1-asp-5freq exercise.

In the last two exercises (t2-gn93 and t6-asp-diff), we test the model dependence of the asteroseismic 
solution, 
and explore which information can actually be obtained on the mixing processes at the convective core boundary. To do 
so, 
we adopted an input physics in the target stars different to that of the grid used for the modelling. 
First, we used the GN93 chemical mixture instead of AGS05 in the t2-gn93 exercise. Then, we included an 
additional diffusive mixing process in the t6-asp-diff case, whilst the grid is computed only with overshooting treated 
as instantaneous mixing.

At the exception of the t6 case, the input frequencies of the t1 and t2 exercises are computed with the non-adiabatic 
MAD code \citep{mad}. This choice is particularly motivated by the t1 case in which same physics than 
the grid is used for the input model. Since the frequencies of the input model could match exactly those of one of the 
grid model, we degrade them by deriving them from a non-adiabatic calculation. This also is more representive of the 
true nature of a real star and avoids a 
possible bias that an ad hoc degradation could generate \citep[see e.g.][about simulating frequencies in 
hare and hound asteroseismic exercises]{reese16}.

\section{Results of the hare and hound exercises}
\label{section5}

We first present the results of the exercises with a common target star (the t1 model), but which differ by the number 
and nature of input constraints.

\subsection{Impact of the seismic dataset}

We selected for this series of tests a model computed with the same physics than in the grid, with the goal of 
isolating 
the only influence of the seismic indicators. Contrary to the grid, the frequencies of this t1 model are computed with 
the MAD non-adiabatic code, as explained in the previous section.

\begin{table*}[!]
\caption{Results of the t1-asp-3freq, t1-asp-4freq, and t1-asp-5freq exercises.}      
 
\label{tablechi2-1}      
\begin{center}                              
\begin{tabular}{l l c c c c c c }        
\hline\hline                 
Exercise & Parameter & \multicolumn{2}{c}{No classical constraint} &  \multicolumn{2}{c}{1-$\sigma$ box 
constraint} &  
\multicolumn{2}{c}{3-$\sigma$ box constraint} \\    
 & & m.f. & MC & m.f. & MC & m.f. & MC \\
\hline        
  \smallskip
  
t1-asp-3freq  & M (14) & 15.6 & 14.6$_{-3.2}^{+2.3}$ & 14 & 14$_{-0.2}^{+0.7}$ & 15.6 & 16.4$_{-1.2}^{+0.9}$\\
\smallskip
  & R  (7.48) & 10.18 & 10.78$_{-3.30}^{+0.97}$ & 7.50 & 7.50$_{-0.04}^{+0.09}$ & 10.18 & 10.49$_{-0.99}^{+1.36}$ \\
\smallskip
  & X (0.70) & 0.70 & 0.70$_{-0.02}^{+0.04}$ & 0.70 & 0.70$_{-0.02}^{+0.04}$ & 0.68 & 0.70$_{-0.02}^{+0.04}$ \\
   \smallskip
  & Z (0.014) & 0.018 & 0.014$_{-0.004}^{+0.004}$ & 0.018 & 0.014$_{-0.002}^{+0.004}$ & 0.018 & 
0.014$_{-0.004}^{+0.002}$ \\
   \smallskip
  & $\alpha_{\mathrm{ov}}$ (0.20) & 0.45 & 0.45$_{-0.15}^{+0.05}$ & 0.10 & 0.20$_{-0.10}^{+0.05}$ & 0.45 & 
0.45$_{-0.20}^{+0.05}$\\      
   \smallskip
 &  $ X_{\mathrm{c}} $ (0.288) & 0.237 & 0.182$_{-0.052}^{+0.069}$ & 0.283 & 0.283$_{-0.011}^{+0.029}$ & 0.237 & 
0.221$_{-0.057}^{+0.036}$ \\
   \smallskip
  & $T_{\mathrm{e}}$ (27647) & 26977 & 25532$_{-3255}^{+2456}$ & 26953 & -- & 26977 & -- \\
   \smallskip
 &  $\log g$ (3.8364) & 3.6150 & 3.5022$_{-0.0696}^{+0.2799}$ & 3.8341 & -- & 3.6150 & -- \\
   \smallskip
 &  $\chi^2$ & 0.0358 & -- & 0.0713 & -- & 0.0358 & -- \\
\hline
                            \smallskip  
  
t1-asp-4freq & M (14) & 13.8 & 13.9$_{-0.2}^{+0.1}$ & 13.8 & 13.8$_{-0.1}^{+0.2}$ & 13.8 & 13.9$_{-0.2}^{+0.1}$ \\  
\smallskip
 & R (7.48) & 7.45 & 7.47$_{-0.02}^{+0.03}$ & 7.45 & 7.45$_{-0.02}^{+0.04}$ & 7.47 & 7.68$_{-0.02}^{+0.03}$\\
\smallskip 
 & X (0.70) & 0.68 & 0.70$_{-0.02}^{+0.02}$ & 0.68 & 0.68$_{-0.02\ddag}^{+0.02}$ & 0.68 & 0.70$_{-0.02}^{+0.02}$ \\
\smallskip 
 & Z (0.014)  & 0.014 & 0.014$_{-0.002\ddag}^{+0.004}$ & 0.014 & 0.014$_{-0.002\ddag}^{+0.004}$ & 0.014 & 
0.014$_{-0.002\ddag}^{+0.004}$ \\
\smallskip 
 & $\alpha_{\mathrm{ov}}$ (0.20)  & 0.20 & 0.20$_{-0.10}^{+0.05\ddag}$ & 0.20 & 0.20$_{-0.10}^{+0.05\ddag}$ & 0.20 & 
0.20$_{-0.10}^{+0.05\ddag}$ \\
\smallskip 
 & $X_{\mathrm{c}} $ (0.288)  & 0.274 & 0.279$_{-0.005}^{+0.027}$ & 0.274 & 0.275$_{-0.001}^{+0.019}$ & 0.274 & 
0.279$_{-0.005}^{+0.027}$ \\
\smallskip 
 & $T_{\mathrm{e}}$ (27647)  & 27888 & 27901$_{-878}^{+476}$ & 27888 & -- & 27888 & -- \\
\smallskip 
 & $\log g$ (3.8364)  & 3.8330 & 3.8330$_{-0.0026}^{+0.0012}$ & 3.8330 & -- & 3.8330 & -- \\
   \smallskip
 &  $\chi^2$ & 0.2272 & -- & 0.2272 & -- & 0.2272 & -- \\
\hline
 \smallskip

t1-asp-5freq  & M (14) & 13.8 & 13.8$_{-0.1}^{+0.1}$ & 13.8 & 13.8$_{-0.1}^{+0.1}$ & 13.8 & 13.8$_{-0.1}^{+0.1}$\\
\smallskip
  & R  (7.48) & 7.45 & 7.46$_{-0.00}^{+0.04}$ & 7.45 & 7.45$_{-0.02}^{+0.02}$ & 7.45 & 7.46$_{-0.00}^{+0.04}$ \\
\smallskip
  & X (0.70) & 0.68 & 0.68$_{-0.02\ddag}^{+0.04}$ & 0.68 & 0.68$_{-0.02\ddag}^{+0.02}$ & 0.68 & 
0.68$_{-0.02\ddag}^{+0.04}$ \\
   \smallskip
  & Z (0.014) & 0.014 & 0.014$_{-0.002\ddag}^{+0.002}$ & 0.014 & 0.014$_{-0.002\ddag}^{+0.002}$ & 0.014 & 
0.014$_{-0.002\ddag}^{+0.002}$ \\
   \smallskip
  & $\alpha_{\mathrm{ov}}$ (0.20) & 0.20 & 0.20$_{-0.05\ddag}^{+0.05\ddag}$ & 0.20 & 0.20$_{-0.05\ddag}^{+0.05\ddag}$ & 
0.20 & 0.20$_{-0.05\ddag}^{+0.05\ddag}$\\      
   \smallskip
 &  $ X_{\mathrm{c}} $ (0.288) & 0.274 & 0.275$_{-0.001}^{+0.031}$ & 0.274 & 0.275$_{-0.001}^{+0.019}$ & 0.274 & 
0.275$_{-0.001}^{+0.031}$ \\
   \smallskip
  & $T_{\mathrm{e}}$ (27647) & 27888 & 27888$_{-1253}^{+100}$ & 27888 & -- & 27888 & -- \\
   \smallskip
 &  $\log g$ (3.8364) & 3.8330 & 3.8330$_{-0.0033}^{+0.0009}$ & 3.8330 & -- & 3.8330 & -- \\
   \smallskip
 &  $\chi^2$ & 0.4175 & -- & 0.4175 & -- & 0.4175 & -- \\
\hline
\end{tabular}

\end{center}
{$\ddag$ indicates when the grid resolution is reached and set as the limit to the confidence interval. 
The m.f. and MC acronyms respectively stand for the analysis based on the input set of observed constraints and 
that from the method with Monte-Carlo simulations (see Sect.~\ref{section2}). The ``no classical constraint`` columns 
give the results without $T_{\textrm{e}}$ and log g used as constraints. The 1-$\sigma$ and 3-$\sigma$ columns gives 
the results when imposing the solutions to be respectively in the 1-$\sigma$ and 3-$\sigma$ boxes on $T_{\textrm{e}}$ 
and log g.  The input stellar parameters of the input model are recalled between brackets in the first column. The 
$M$, $R$, $T_{\mathrm{e}}$, and $g$ parameters are given in M$_{\odot}$, R$_{\odot}$, K, and cm/s$^2$, respectively.}
\end{table*}

\subsubsection{Three frequencies with known angular degree: The t1-asp-3freq test}
\label{section4-1}
\begin{figure}
\centering
\includegraphics[width=9.4cm]{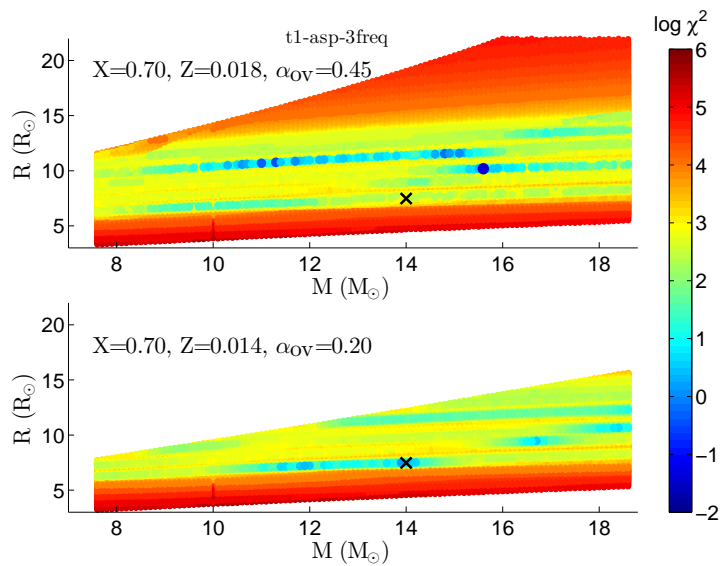}
      \caption{Values of $\log \chi^2$ for the input frequencies of the t1-asp-3freq 
exercise are represented with a colour map (see legend in the figure for the related scale) in the $M$--$R$ plane of 
the 
grid of models. The chemical composition and overshooting ($X, Z, \alpha_{\textrm{ov}}$) are those of the 
$\chi^2$ global minimum in the top panel and those of the input model t1 in the bottom panel. A cross symbol is 
located at the real position of the t1 model.}
         \label{Fig1-MR-3freq}
   \end{figure}

The set of frequencies for this exercise is composed of one mode of each degree\footnote{For 
clarity, hereafter, we use the notations $\ell_i$ and $k_i$ to refer to a mode of degree $\ell=i$ and radial 
order $k=i$.} $\ell$=0 to 2 (see 
Table~\ref{table2-section4}). The solutions obtained without classical constraints and on the sole seismic 
dataset as input are given in the 3rd column (No classical - m.f.) of Table~\ref{tablechi2-1}. It predicts for 
$M$, $R$, and $\alpha_{\textrm{ov}}$ values of 15.6 $\msun$, 10.18~$\rsun$, and 0.45, respectively. This fails at 
reproducing the true stellar parameter of the t1 model, and leads to a clear overestimation of 
overshooting.  Figure~$\ref{Fig1-MR-3freq}$ shows the map of the $\chi^2$ function in the  $M$--$R$ 
plane. For illustrating the discrepancy, the chemical composition and $\alpha_{\mathrm{ov}}$ in the top panel are set 
to those of the 
solution: the global minimum clearly does not lie close to the input stellar parameters that were to be retrieved. The 
map also reveals regions with lower $\chi^2$ values under the form of ridges (blue patterns). They correspond 
to places of same 
$\tau_{\textrm{dyn}}$.  This is expected due to the presence of a radial mode in the input set and the direct 
dependence of 
radial modes to $\tau_{\textrm{dyn}}$. However we see in the bottom panel of Fig.~\ref{Fig1-MR-3freq} that a local 
minimum is present at the correct $M$--$R$ location when the 
chemical composition and $\alpha_{\mathrm{ov}}$ of the t1 input model are adopted. 

The solution obtained with the re-sampling method, presented in Sect.~\ref{section2}, is given in the 4th column (No 
classical - MC) of Table~\ref{tablechi2-1}, still without including classical constraints. The inferred mass, with a 
value of 
14.6~$\msun$, is now close to that of the t1 input model (14~$\msun$) but the 1-$\sigma$ error signals a low 
precision, $\sim $22\%. The radius and the overshooting parameter are still significantly overestimated. We explain 
this trend in more detail hereafter. 

\bfseries Including classical constraints. \mdseries When we require the solution to satisfy within 1-$\sigma$ the 
$T_{\textrm{e}}$ and $\log g$ classical constraints, most of the stellar 
parameters of the input model are now correctly retrieved 
(see column 5 of Table~\ref{tablechi2-1}), at the exceptions of $\alpha_{\textrm{ov}}$ and $Z$, which are 
underestimated by 0.10 and overestimated by 0.004, respectively. The determination of these two parameters actually 
presents a degeneracy. Indeed, the overshooting extends the main-sequence lifetime and also increases the luminosity of 
the evolutionary track. On the other hand, the chemical composition in the stellar 
envelope determines the opacity, hence the escaping radiation and luminosity: for a given mass, the lower the 
metallicity is, the higher the luminosity is. Stellar models with a given mass but with different combinations of 
overshooting and chemical 
composition can thus correspond to a same luminosity.

\begin{figure}
\centering
\includegraphics[width=9.4cm]{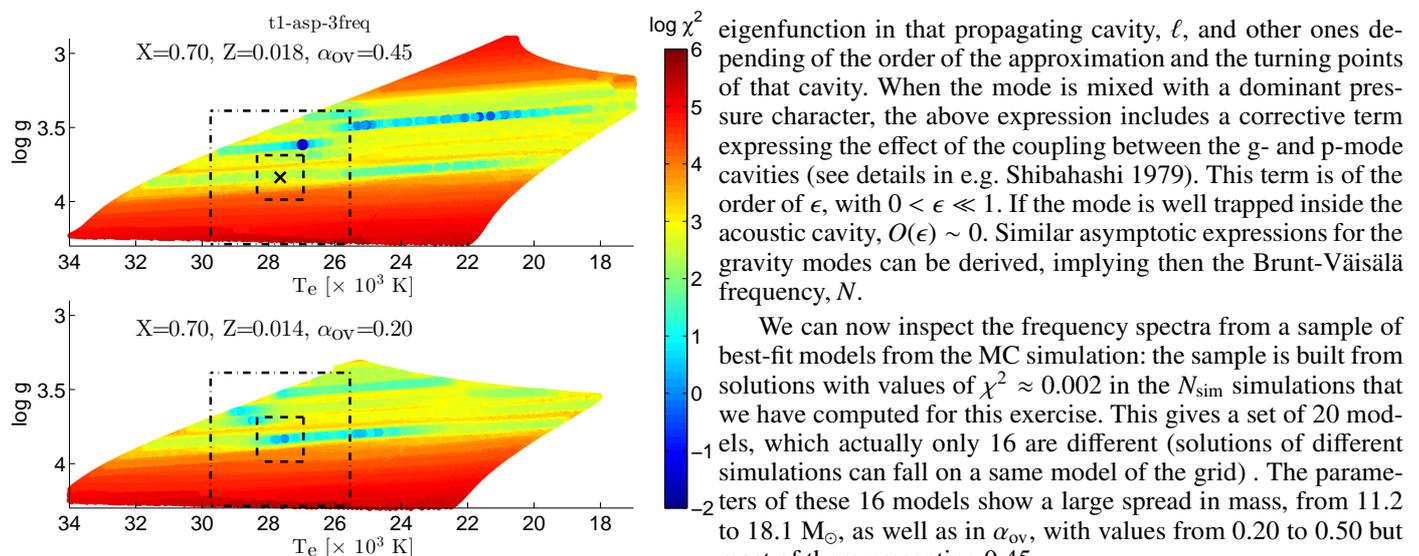}
      \caption{Same as in Fig.~\ref{Fig1-MR-3freq}, but in the $T_{\textrm{e}}$--$\log g$ plane.}
\label{Fig1-Tefflogg-3freq}
\end{figure}  

Figure~\ref{Fig1-Tefflogg-3freq} shows the $\chi^2$ map now in the  $T_{\textrm{e}}$--$\log g$ plane. It illustrates 
how 
the addition of 
the fundamental parameters as constraints discard the global minimum, which is outside the 1-$\sigma$ box 
on $T_{\textrm{e}}$ and $\log g$. However, despite the 
presence of a local 
minimum at the exact location of the target (bottom panel), the solution that is obtained remains inaccurate on some 
stellar parameters, as we 
have 
seen above. If we relax the precision on the classical parameters by considering a 3-$\sigma$ error box, then the 
global minimum lies within that error box. Consequently, 
considering a too loose constraint (3-$\sigma$ error box), or dealing with larger errors on classical parameters, 
which 
can be specially the 
case for $\log g$, do not help improve the solution in 
comparison to that solely based on the seismic constraints.

With the inclusion of the classical parameters, the re-sampling method appears more robust. The 
inferred parameters then match those of the input model (see column 6 of Table~\ref{tablechi2-1}). The errors at the 
1-$\sigma$ confidence level are down to 5\% on the mass 
and $\sim$1.3\% on the radius. Meanwhile the overshooting and metallicity are also correctly predicted, with an 
estimated precision of $\Delta Z$=0.004 and $\Delta \alpha_{\textrm{ov}}$~=~0.10. Again, the 3-$\sigma$ 
classical constraints do not change the results of the MC simulations, providing the same results than without using 
the classical constraints. 

\bfseries The high-overshooting bias. \mdseries To understand the tendency of the solution based on 
the only seismic dataset to favour high values of overshooting, we investigated in more detail the frequency spectra 
of the solutions emerging from the MC solutions. Before, it is useful to characterise some properties of the 
mixed modes.  
Firstly, the ratio of the 
vertical kinetic energy, $E_{k,v}$, to the total kinetic energy, $E_{k}$, reflects the dominant g- or p-character of 
an oscillation mode. The ratio is close to 0 for pure gravity modes, while it is equal to 1 for radial modes 
and typically greater than $\sim$~0.9 for non-radial pressure modes. In the case of mixed modes, this ratio 
ranges in 
between these values.

Asymptotic developments allow to estimate the frequencies of pressure or gravity modes with help of simple 
relations based on the structural quantities of the stellar models \citep[e.g.][]{shiba79,tassoul}. In the case of 
low-degree pressure modes as in $\beta$ Cephei stars, these developments need in principle to include higher-order 
terms 
\citep[][]{tassoul90,roxburgh94,smeyers96}. Despite these levels of refinement, the asymptotic 
approach does not formally apply to low-radial order modes. However, it remains useful to estimate frequency values 
and relate them to the stellar structure properties, while it should not be used to compute the frequencies in a 
precise asteroseismic modelling of a $\beta$ Cephei star. An asymptotic expression of frequencies for acoustic 
modes reads as:
\begin{equation}
\sigma \simeq A \left[\int_{r_{p_{\ell}}}^{R} \frac{1}{c_s} dr\right]^{-1}
\label{eqchi2-2}
\end{equation}
with $c_s$ the local sound speed, $r_{p_{\ell}}$ the radius defining the lower limit of the propagating cavity of 
the pressure mode, and $A$ a sum function wich includes as terms the number of nodes of the 
eigenfunction in that propagating cavity, $\ell$, and other ones depending of the order of the 
approximation and the turning points of that cavity. When the mode is mixed with a dominant pressure 
character, the above expression includes a corrective term  expressing the effect of the coupling between the g- and 
p-mode cavities \citep[see details in e.g.][]{shiba79}. This 
term is of the order of $\epsilon$, with 0~$<\epsilon\ll$~1. If the mode is well trapped inside the 
acoustic cavity, $O(\epsilon)\sim 0$.  Similar asymptotic expressions for the gravity modes can be derived,
implying then the Brunt-V\"ais\"al\"a frequency, $N$.

We can now inspect the frequency spectra from a sample of best-fit models from the MC simulation: the sample is built 
from solutions with values of $\chi^2\approx$~0.002 in the $N_{\textrm{sim}}$ simulations that we have computed for 
this 
exercise. This gives a set of 20 models, which actually only 16 are different (solutions of different simulations can 
fall on a same model of the grid) . The parameters of these 16 models show a 
large spread in mass, from 11.2 to 18.1~$\msun$, as well as in $\alpha_{\textrm{ov}}$, with values from 0.20 to 0.50 
but most of them presenting 0.45.

\begin{figure*}[!]
\begin{center}
\includegraphics[width=0.45\textwidth]{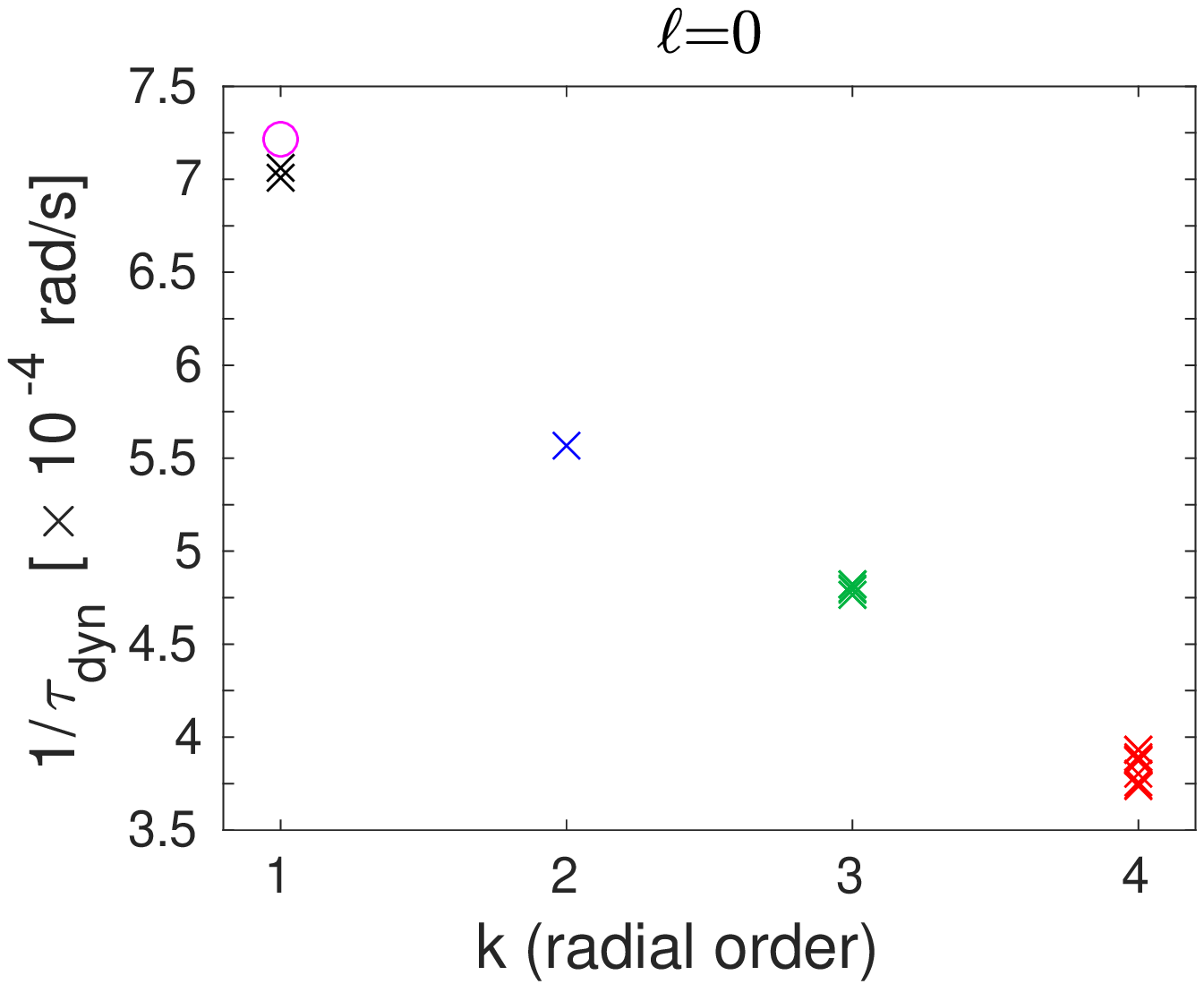} 
\includegraphics[width=0.45\textwidth]{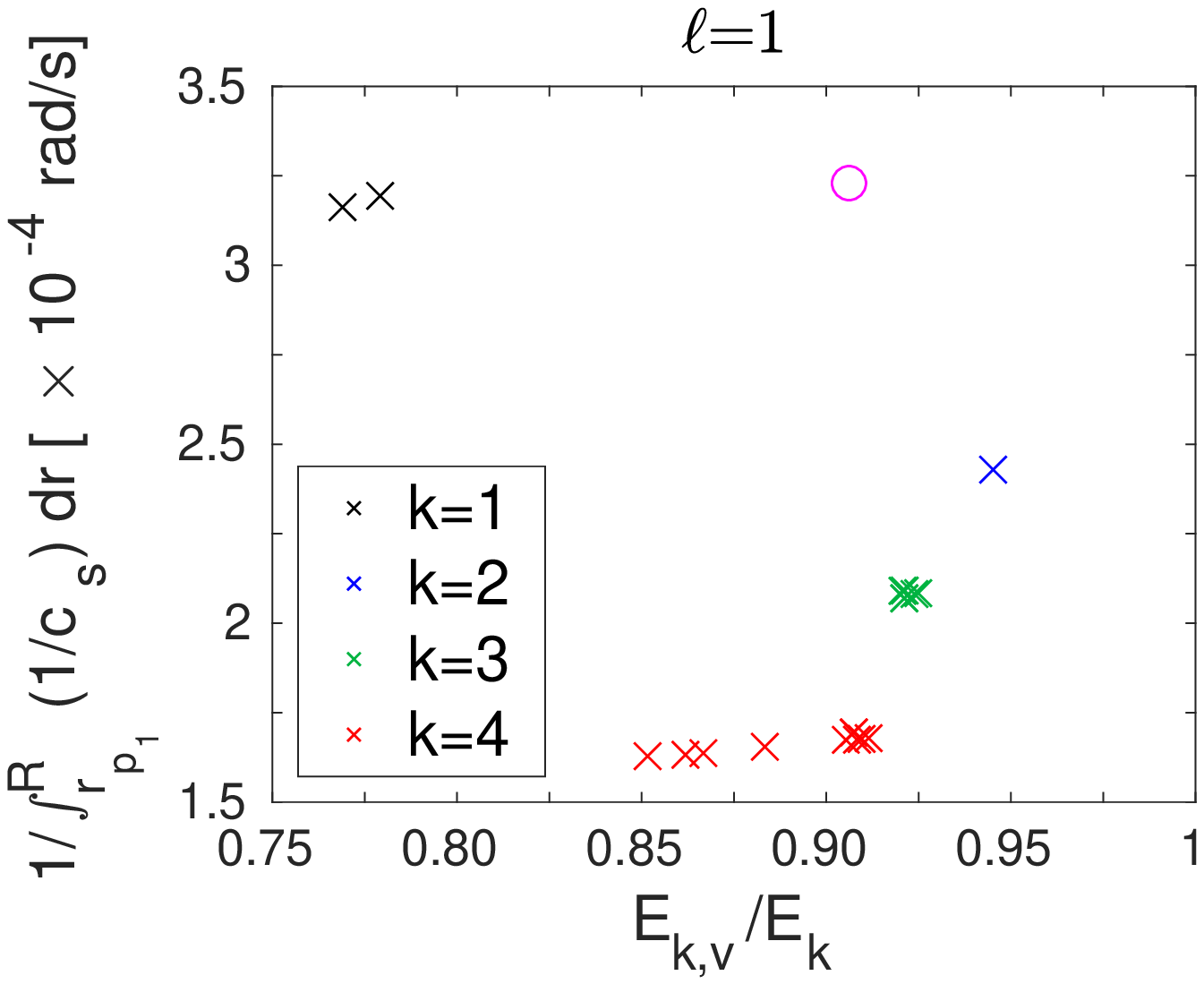}
\includegraphics[width=0.45\textwidth]{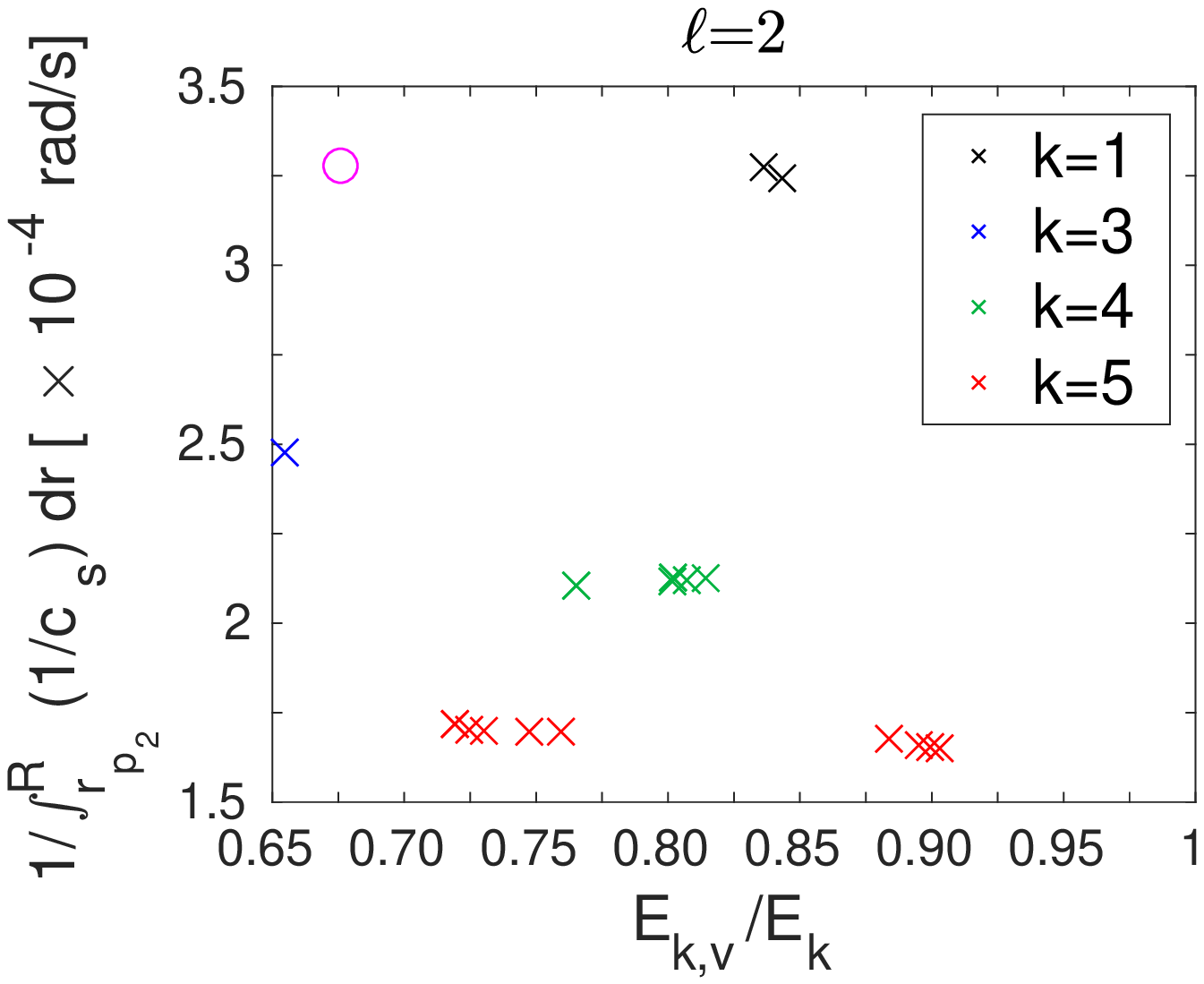}
\end{center}
\caption{Results of the MC simulations for the t1-asp-3freq test. We select a sample of 20 models presenting similar 
values of $\chi^2$ as the global minimum. Top left panel: inverse of the dynamical timescale as a function of the 
radial order of the $\ell_0$ modes (in the 20 
models described above) that best fit the original $\ell_0$ mode of the t1-asp-3freq input frequency set. Top 
right panel: $[\int_{r_{p_1}}^{R} (1/c_s) dr ]^{-1}$ as a function of $E_{k,v}/E_{k}$ for the $\ell_1$ modes that 
best fit the original $\ell_1$ mode of the t1-asp-3freq frequency in the selected models. The colour index 
represents the radial order of the fitting $\ell_1$ modes. Bottom panel: same as in the top right panel, but now 
adapted for the $\ell_2$ modes. The case of the input t1 model is represented in each panel by an open magenta circle.}
\label{figchi2-4}
\end{figure*}

We compare in Fig.~\ref{figchi2-4} the values of $1/\tau_{\tau_{\mathrm{dyn}}}$ and the integral of 
Eq.~(\ref{eqchi2-2}) for the radial and non-radial 
modes in the 16 models that were best fitting the input frequencies. We can 
identify four groups. Black crosses correspond to the solutions in which the three fitting modes are: ($\ell_0, 
k_1$), ($\ell_1, k_{1}$) and ($\ell_2, 
k_{1}$) modes. These radial orders and angular degrees are exactly the same as those of the t1-asp-3freq frequency set 
(magenta circles). The other three groups are composed of [$\ell_0, k_i$; $\ell_1, k_{i}$; $\ell_2, k_{i+1}$] modes, 
with $i$=2 (blue crosses), 3 (green crosses), and 4 (red crosses). 

The models of the first group, as they reproduce exactly the $k$ order of the radial mode, have a 
similar $\tau_{\textrm{dyn}}$ than the t1 model. In the three other groups of models, each of the input frequencies is 
actually fitted by a mode of a higher overtone. The solutions with a radial 
mode presenting a higher $k_i$ present larger $\tau_{\textrm{dyn}}$ than the t1 model, 
since the proportionality of radial mode frequencies to $\tau_{\textrm{dyn}}$ includes a factor $k$. We see in the top 
left panel that the 
solutions indeed gather around similar values of 
$\tau_{\textrm{dyn}}$ according to the radial order of their matching $\ell_0$ mode.  The non-radial modes also gather 
according to their radial order (see top right and bottom panels). Two modes from two different stellar structures and 
with different radial orders will be characterised by values $A$ and $A'$ in Eq.~(\ref{eqchi2-2}). The modes then will 
have a similar frequency value if the ratio of the integrals appearing in Eq.~(\ref{eqchi2-2}) is proportional to 
$A/A'$. Thus, the non-radial modes in 
Fig.~\ref{figchi2-4} group 
according to $k$ because they have to reproduce a similar value of the ratio to present a frequency similar to, and 
hence fitting, that of the t1 model. As $k$ increases, so does $A$, because it is $\propto k$. Consequently the 
inverses of the integrals of the modes with higher radial orders than the t1 case have to present lower values than the 
t1 inverse of the integral, as it is indeed 
observed in Fig.~\ref{figchi2-4}.

Overshooting now explains how models with different parameters can succeed in reproducing similar $\tau_{\textrm{dyn}}$ 
and inverse integrals of the sound speed. Most of the selected models present indeed higher overshooting values than 
the t1 
model. 

As we have seen, provided $\tau_{\textrm{dyn}}$ is higher than in the t1 model, higher overtone radial modes can 
match the frequency of the t1 fundamental radial mode. For models with large overshoot mixing, the range of 
values reached by their radius during the main sequence is increased. Since $\tau_{\textrm{dyn}}^2\propto R^3M^{-1}$, 
these models will also evolve through a larger range of 
$\tau_{\textrm{dyn}}$. This increases the probability that during their MS, a higher overtone in their 
frequency spectra matches that of the input radial mode, even if their fundamental radial mode would never present the 
same frequency than the t1 one.

Similarly, the increased variety of internal sound speed profiles and radii of a high-overshooting model will increase 
the range of values of their $\int_{r_{p_{\ell}}}^{R} 1/c_s dr$ integrals for $\ell_1$ and $\ell_2$ modes. Therefore 
the probability also increases to reach a value of the integral, which the ratio with the t1 integral 
allows higher-overtones modes to reproduce the frequency of the lower-overtone t1 modes.

The use of the classical parameters has prevented this bias because they put an 
additional constraint on the radius and discarded models which did not reproduce  $\tau_{\textrm{dyn}}$. It is 
thus crucial to determine them independently with the highest accuracy and precision possible when a limited number 
of modes are considered for the asteroseismic modelling. Recently, the first interferometric measurement of the 
angular diameter of a $\beta$ Cephei star ($\beta$ Can Maj) by \citet{veritas20} has reached a precision of $\sim 
3$\%. In 
combination with Gaia parallaxes, 
the interferometric measurement of radii of $\beta$ Cephei stars would be a promising 
additional classical constraint on which to rely.

We also performed additional tests (not presented in detail here), first where a second $\ell_1$ mode instead of the 
$\ell_2$ one was considered in the input dataset. In that case, the results of the seismic modelling and the 
re-sampling 
method did not change for the mass, but gave accurate results for $R$. The $\alpha_{\textrm{ov}}$ parameter 
overestimation was 
reduced ($\alpha_{\textrm{ov}}=0.30$), and with the re-sampling method, it reduced to 
$\alpha_{\textrm{ov}}=0.25^{+0.15}_{-0.10}$. The 
precision was not 
improved. The frequency spacing between mixed modes of same angular degree brings additional information on the 
structure, and reduces the probability to match frequency spectra by higher overtones of a model with a totally 
different structure. However, in another test where the three modes where all $\ell_1$, the solution was significantly 
degraded. It appears clearly it came from a degeneracy on the determination on $M$ and $R$ in the absence on a 
constraint on $\tau_{\textrm{dyn}}$ through a radial mode. 

In conclusion, the ranges of global stellar properties able to lead to a good fit of a small number of frequencies are 
quite large. 
Fitting these frequencies then do not allow to characterise reliably the input model, without help of additional 
constraints.
   
\subsubsection{Four frequencies with known angular degree: The t1-asp-4freq test}   

The input set of frequencies is composed of one radial mode and three non-radial modes (2 $\ell_1$, 1 $\ell_2$). The 
results are reported in Table~\ref{tablechi2-1}: $Z$ (0.014) and 
$\alpha_{\textrm{ov}}$ 
(0.20) are perfectly fitted, while the other stellar parameters are slightly underestimated with 
$M$~=~13.8~M$_{\odot}$, 
$R$~=~7.45~$R_{\odot}$, $X$~=~0.68, and $X_{\textrm{c}}$~=~0.274.

In comparison with the tests composed of three frequencies, we checked the $\chi^2$ map and found it contains fewer 
local 
minima. The global minimum lies now very close to the actual location of the t1 input model. With the re-sampling 
method, the inferred stellar parameters are then even closer to the t1 model (column 4 of Table~\ref{tablechi2-1}). 
This method accounts for the theoretical uncertainties, which we have seen in Sect.~\ref{section2-3} can be dominated 
by non-adiabatic effects. Here, as the target model and the model grid have the same physics, the non-adiabaticity of 
the input frequencies seems likely at the origin of the small inaccuracy in some of the retrieved stellar parameters.

\bfseries Including classical constraints. \mdseries With or without these constraints, the errors are 
very low in the three cases (columns 4,6, and 8 of Table~\ref{tablechi2-1}), reaching a precision of $\sim$1\% on $M$ 
and 
$\sim$0.005\% on $R$. As observed in the additional tests with three frequencies, the 
seismic modelling when it includes two mixed modes of same $\ell$ degree benefits from an information on 
the evanescent region. This region is a strong marker of the stellar structure as it is defined by the layers 
marking the transition between the core and radiative envelope.

\subsubsection{Five frequencies with known angular degree: The t1-asp-5freq test}   

In this exercise, the number of seismic constraints is equal to the degrees of freedom of the problem, and would begin 
in 
principle to be optimal for adjusting the free parameters. The input frequency set is composed of the same frequencies 
than the 
t4-asp-op-4freq exercise, but includes one additional $\ell_2$ mode. 

\begin{figure}
\centering
\includegraphics[width=9.4cm]{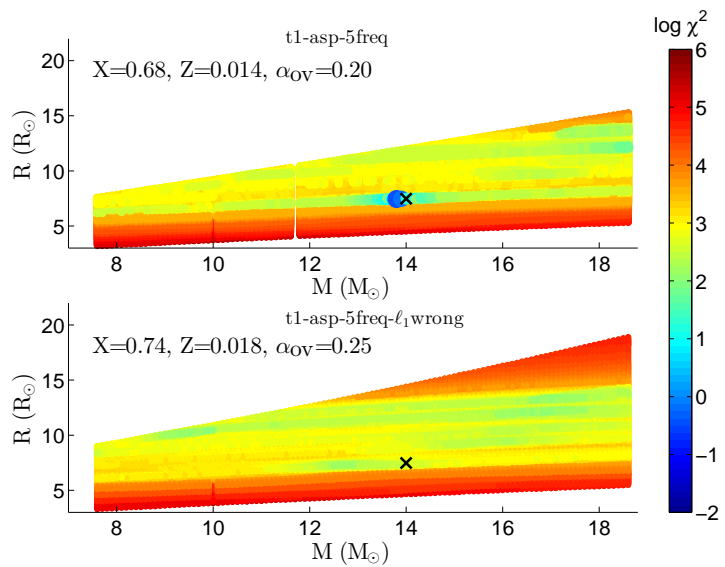}
      \caption{Same as in Fig.~\ref{Fig1-MR-3freq}, but for the t1-asp-5freq and t1-asp-5freq-$\ell_1$wrong cases 
in the top and bottom panels, respectively.}
         \label{Fig2-MR-5freq}
   \end{figure}

The model minimising $\chi^2$ appears to correspond to the same model that in the preceding exercise with only four 
frequencies, so that the inferred stellar parameters are identical. We observe similarly a significant 
decrease of the number of local 
minima and their $\chi^2$ value, which is illustrated in the top panel of Fig.~\ref{Fig2-MR-5freq}. In comparison to 
the previous exercise, we observe an increased precision on the determination of the chemical composition and on 
$\alpha_{\textrm{ov}}$, which reaches in this case the limit in precision of the grid. The stellar parameter 
distributions from the solutions of the MC simulations show that $>$80\% of the models are 
characterised by $\alpha_{\textrm{ov}}$~=~0.20 and $\sim$70\% of them 
present a mass of 13.8 or 13.9~M$_{\odot}$.   

An inaccuracy of 0.2 $\msun$ appears on the mass derived by the seismic modelling and is very 
likely related to the non-adiabatic computation of the input frequencies. It also reveals that the values we 
estimated 
for the errors could be underestimated, since the mass of the target is not predicted by our 1-$\sigma$ interval of 
confidence:  its upper limit predicts a mass of 13.9 $\msun$, however only at 0.1 $\msun$ of the real value.

         \begin{figure}
\centering
\includegraphics[width=0.48\textwidth]{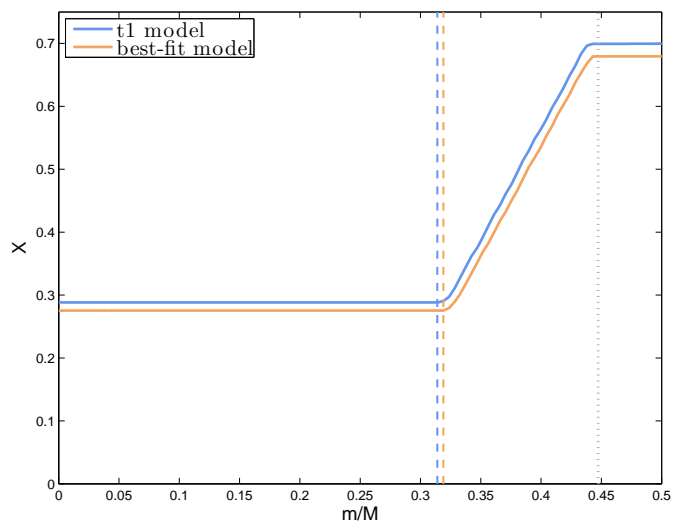}
      \caption{Profile of the hydrogen abundance, X, as a function of the normalised mass, for the target t1 model in 
blue and the best-fit model of the t1-asp-5freq (and t1-asp-4freq) exercise in orange. The vertical dashed lines 
indicate the limits of the fully mixed central region (i.e. including the overshoot region) of the two models, with the 
same colour association than the X profile. The vertical dotted lines are similarly colour-coded and indicate the 
limits of the gradient of chemical composition.}
         \label{Fig1-mm-5freq}
   \end{figure}   

The present asteroseismic modelling has shown its ability to retrieve the global stellar parameters with high 
precision and accuracy. But we also find that it is able to determine the location of the central fully 
mixed 
region -the convective core and the overshooting region- both in terms of the radius and more importantly for stellar 
evolution, of mass with a high accuracy (see values in Table~\ref{tablechi2-11}). In 
Fig.~\ref{Fig1-mm-5freq} the locations in term of the normalised mass, $m/M$, of this region in 
the t1 target model and in the best-fit model match to less than 1\%. 
Similarly, the gradient of chemical composition ($\nabla \mu$ hereafter) above the boundary of the mixed 
region is also reproduced with an accuracy~<1\%. The shape and extent of this gradient are a marker of 
the extra-mixing processes. Succeeding in delivering a tight constraint on it is key to reveal the nature of these 
processes in massive stars.
   
\begin{table*}[!]
\caption{Boundaries of the fully-mixed central region in terms of radius($r_{cc}$) or mass ($m_{cc}$) and $\nabla \mu$ 
region in terms of radius ($r_{\mu}$) or mass ($m_{\mu}$) for targets and their best-fit models in different 
exercises.}             
\label{tablechi2-11}      
\begin{center}
\begin{tabular}{l c c c c c c c c c c c c}        
\hline\hline                 
Exercise & $r^i_{cc}$& $r^i_{\mu}$ & $r^o_{cc}$ & $r^o_{\mu}$  & $\Delta r_{cc}/r_{cc}$ & $\Delta r_{\mu}/r_{\mu}$  & 
$m^i_{cc}$ & $m^o_{cc}$ & $m^i_{\mu}$ & $m^o_{\mu}$ & $\Delta m_{cc}/m_{cc}$ & $\Delta m_{\mu}/m_{\mu}$\\   
 &  $(r/R)$ & $(r/R)$ & $(r/R)$ &  $(r/R)$ &  &   &  $(m/M)$ & $(m/M)$ & $(m/M)$ & $(m/M)$ & &  \\   
\hline  
t1-asp-op-4freq & 0.153 & 0.194 & 0.154 & 0.193 & -0.007 & -0.005 & 0.315 & 0.319 & 0.448 & 0.448 & -0.01 & 0.00 \\     
t1-asp-op-5freq & & & & & & \\  
\hline
t2-gn93 & 0.159 & 0.195 & 0.169 & 0.204 & -0.06 & -0.05 & 0.291 & 0.337 & 0.408 & 0.457 & -0.16 & -0.12\\       
\hline  
t6-asp-diff & 0.152 & 0.233 & 0.158 & 0.191 & -0.04 & 0.18 & 0.237 & 0.222 & 0.471 & 0.337 & 0.06 & 0.28\\
\hline                                   
\end{tabular}
 \end{center}
Note: The variables indexed by $i$ and $o$ respectively represent those of the input target model and the 
output best-fit model from the exercise indicated in the first column.
\end{table*}   
   
As expected, the classical constraints do not improve further the modelling results and their errors. However, since 
the classical parameters are extracted from the t1 model, they do not suffer inaccuracy as could parameters derived 
from 
real conditions observations. The approach could thus be reversed in the study of observed $\beta$ Cephei stars 
by comparing whether parameters such as $T_{\textrm{e}}$ and $\log g$ seismically determined  match their 
photometric or spectroscopic determinations. In particular, a clear discrepancy between the asteroseismic and 
spectroscopic $\log g$ was revealed by the modelling of several $\beta$ Cephei stars, as discussed in 
\citet{aerts11}. These authors suggest the origin of the disagreement might be due to the pulsational broadening of 
lines not well accounted for in the spectroscopic analysis used for deriving the surface gravity.

\subsection{Knowledge of the mode identification} 

We first analysed the impact of misidentifying a $\ell$ degree. In 
practice, although the methods\footnote{such as the moment method based on spectroscopic line variations 
\citep{balona86,aerts96} or the analysis of photometric bandpass ratio \cite[][]{cugier94}.} developed for 
identifying the modes in $\beta$ Cephei stars are generally able to constrain the angular degree of detected 
pulsations, in some cases the identification can be ambiguous. That was for example the case of $\nu$ Eri 
\citep[see][in particular their Figs.~5 and 6]{deridder}, whatever the method used. The same difficulty occurred for 
the 
photometric bandpass ratios analysis of the 12~Lac star \citep{handler06}. In a second time, we have focused on cases 
where 
no identification of the mode is possible, which is typically the case of stars observed in a single photometric 
bandpass.

\subsubsection{Misidentifying a mode}

We assume in the t1-$\ell$1wrong exercise the same set of five frequencies as in the t1-asp-5freq case, but one of the 
$\ell_1$ modes is misidentified as an $\ell_2$. The inferred 
parameters clearly fail at reproducing the t1 model: for instance, they give $M$~=~10.10~M$_{\odot}$, 
$R$~=~10.45~R$_{\odot}$, and $X_{\textrm{c}}$~=~0.004. As illustrated in the bottom panel of 
Fig.~\ref{Fig2-MR-5freq}, the $\chi^2$ map degrades in comparison to the case where frequencies were correctly 
identified 
(top panel). The $\chi^2$ global minimum value signals the issue: it increases to $\chi^2$~=~85.162, about 200 to 400 
times the corresponding values in the t1-asp-5freq and t1-asp-4freq exercises, respectively.

\begin{figure}
\begin{center}
\includegraphics[width=0.48\textwidth]{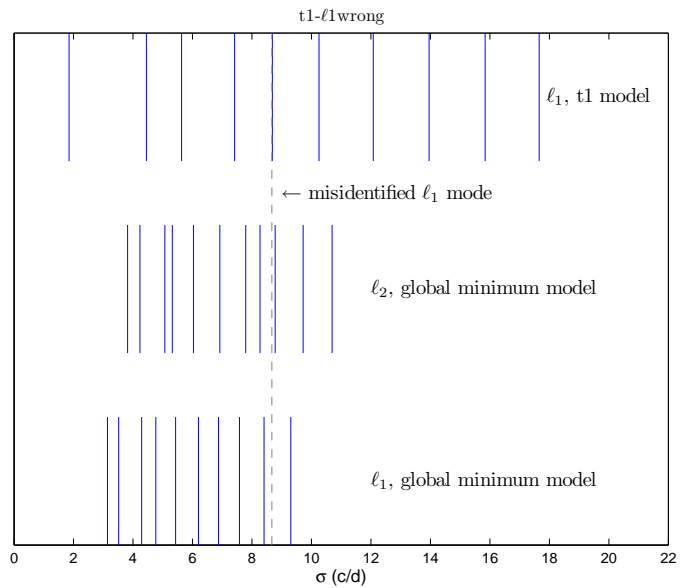}
\end{center}
\caption{Frequency spectra of the t1 model and that of the best-fit model of the t1-$\ell$1wrong test. The frequency of 
the erroneous mode is represented by a dashed grey line. The $\ell$ degree of the different frequency 
spectra is indicated to the right.}
\label{figchi2-11}
\end{figure}

The best-fit model is evolved and close to the TAMS. Its frequency spectrum is denser than in a less-evolved 
stellar model. Fig.~\ref{figchi2-11}  depicts the frequency spectra of the 
$\ell_1$ and $\ell_2$ modes of this model, as well as the $\ell_1$ spectrum of the model t1. The grey dashed 
line indicates the $\ell_1$ input mode that is identified as an $\ell_2$ one. With the misidentification, the 
frequency spacing between the input $\ell_2$ modes is erroneously reduced. Since the input modes are associated 
to theoretical 
mode with the same $\ell$ degree, the 
solution is directed to models reproducing that erroneous feature, that is with denser frequency spectra, in a 
later stage of evolution.

\bfseries Including the classical constraints. \mdseries When imposing the solution to fall within the 1-$\sigma$ error 
box on $T_{\textrm{e}} - \log g$, the inferred parameters are improved; they predict $M$~= 
14.2~M$_{\odot}$, $R$~=~7.77~R$_{\odot}$, and $X_{\textrm{c}}$~=~0.226, close to the t1 input ones. 
However, $\alpha_{\textrm{ov}}$~=~0 and a high $Z$ of 0.018 appear in contradiction with the t1 parameters. The 
minimum, $\chi^2$=554.05, again signals an issue with the result of the modelling.

\subsubsection{Unknown identification of the modes}

We determined the consequences of having no information on the angular degree through the t1-$\ell$-4unknown and
t1-$\ell$-5unknown exercises. We took the same four and five frequencies than in the t1-asp-4freq and 
t1-asp-5freq 
tests, respectively, but without knowledge of their $\ell$ degrees. The results of the two exercises are given in 
Table~\ref{tablechi2-2}.

\begin{table*}[!]
\caption{Results of the t1-asp-4freq-undef and t1-asp-5freq-undef exercises.}             
\label{tablechi2-2}      
\begin{center}                       
\begin{tabular}{l l c c c c c c }        
\hline\hline                 
Exercise & Parameter & \multicolumn{2}{c}{No constraint} &  \multicolumn{2}{c}{1-$\sigma$ box constraint} &  
\multicolumn{2}{c}{3-$\sigma$ box constraint} \\  
 & & m.f. & MC & m.f. & MC & m.f. & MC \\
\hline                   
\smallskip
t1-asp-4freq-undef & M (14) & 14.2 & 14.2$_{-2.8}^{+2.8}$ & 16.6 & 16.6$_{-0.4}^{+0.4}$ & 16.6 & 
16.7$_{-2.1}^{+1.0}$ 
\\  
\smallskip
 & R (7.48) & 9.99 & 10.02$_{-0.55}^{+0.99}$ & 9.43 & 9.43$_{-0.07}^{+0.09}$ & 9.43 & 9.64$_{-0.18}^{+0.66}$\\
\smallskip 
 & X (0.70) & 0.72 & 0.72$_{-0.04}^{+0.02}$ & 0.70 & 0.70$_{-0.02\ddag}^{+0.04}$ & 0.70 & 0.72$_{-0.04}^{+0.02}$ \\
\smallskip 
 & Z (0.014)  & 0.016 & 0.014$_{-0.002}^{+0.004}$ & 0.016 & 0.016$_{-0.002}^{+0.002}$ & 0.016 & 
0.014$_{-0.002}^{+0.002}$ \\
\smallskip 
 & $\alpha_{\mathrm{ov}}$ (0.20)  & 0.35 & 0.35$_{-0.20}^{+0.10}$ & 0.15 & 0.15$_{-0.15}^{+0.10}$ & 0.15 & 
0.25$_{-0.15}^{+0.15}$ \\
\smallskip 
 & $X_{\mathrm{c}} $ (0.288)  & 0.201 & 0.177$_{-0.115}^{+0.078}$ & 0.212 & 0.212$_{-0.035}^{+0.052}$ & 0.212 & 
0.212$_{-0.074}^{+0.072}$ \\
\smallskip 
 & $T_{\mathrm{e}}$ (27647)  & 25197 & 25197$_{-3379}^{+3238}$ & 28084 & -- & 28084 & -- \\
\smallskip 
 & $\log g$ (3.8364)  & 3.5909 & 3.5909$_{-0.1530}^{+0.1257}$ & 3.7087 & -- & 3.7087 & -- \\
   \smallskip
 &  $\chi^2$ & 0.0269 & -- & 0.0427 & -- & 0.0427 & -- \\
  \smallskip
t1-asp-5freq-undef & M (14) & 15 & 14.4$_{-3.1}^{+2.0}$ & 13.8 & 13.8$_{-0.2}^{+0.1}$ & 15 & 15$_{-1.0}^{+1.8}$ \\  
\smallskip
 & R (7.48) & 10.16 & 10.04$_{-2.30}^{+1.00}$ & 7.45 & 7.45$_{-0.03}^{+0.02}$ & 10.16 & 10.10$_{-1.09}^{+0.94}$\\
\smallskip 
 & X (0.70) & 0.74 & 0.72$_{-0.04}^{+0.02}$ & 0.68 & 0.68$_{-0.02\ddag}^{+0.02}$ & 0.74 & 0.72$_{-0.04}^{+0.02}$ \\
\smallskip 
 & Z (0.014)  & 0.014 & 0.014$_{-0.004}^{+0.002}$ & 0.014 & 0.014$_{-0.002\ddag}^{+0.002}$ & 0.014 & 
0.014$_{-0.004}^{+0.002}$ \\
\smallskip 
 & $\alpha_{\mathrm{ov}}$ (0.20)  & 0.35 & 0.35$_{-0.15}^{+0.05}$ & 0.20 & 0.20$_{-0.05\ddag}^{+0.05\ddag}$ & 0.35 & 
0.35$_{-0.15}^{+0.05}$ \\
\smallskip 
 & $X_{\mathrm{c}} $ (0.288)  & 0.210 & 0.200$_{-0.188}^{+0.073}$ & 0.274 & 0.275$_{-0.001}^{+0.019}$ & 0.210 & 
0.201$_{-0.081}^{+0.074}$ \\
\smallskip 
 & $T_{\mathrm{e}}$ (27647)  & 25699 & 25713$_{-3556}^{+1983}$ & 27888 & -- & 27888 & -- \\
\smallskip 
 & $\log g$ (3.8364)  & 3.6001 & 3.5931$_{-0.0538}^{+0.1322}$ & 3.8330 & -- & 3.6001 & -- \\
   \smallskip
 &  $\chi^2$ & 0.2452 & -- & 0.4175 & -- & 0.2452 & -- \\
\hline
\end{tabular}
\end{center}
{Same comments than in Table \ref{tablechi2-1}.}
\end{table*}

At first, with a set of four frequencies, the mass deduced is not too far away from the input one, but the radius and 
overshooting are overestimated to $R$~=~9.99~R$_{\odot}$ and $\alpha_{\textrm{ov}}$~=~0.35, as reported in column 3 
of Table~\ref{tablechi2-2}.

\begin{figure}
\centering
\includegraphics[width=0.45\textwidth]{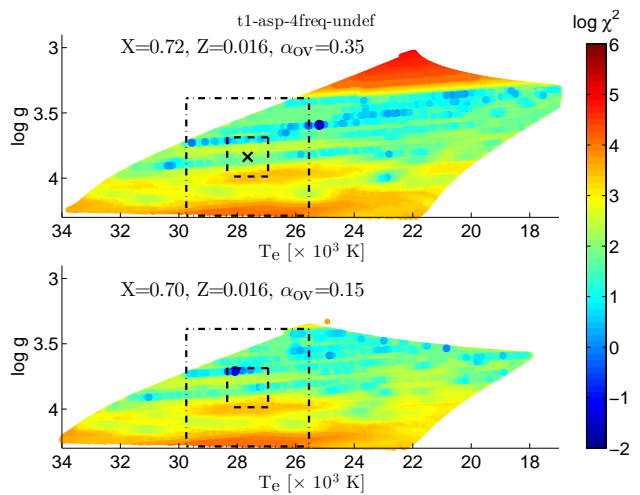}
      \caption{Same as in Fig.~\ref{Fig1-Tefflogg-3freq}, but for the t1-asp-4freq-undef case.}
         \label{Fig3-Tefflogg-4freq}
   \end{figure}

In the top panel of Fig.~\ref{Fig3-Tefflogg-4freq}, the $\chi^2$ map in the $T_{\textrm{e}}$--$\log g$ plane for the 
output 
parameters clearly 
illustrates that the global minimum predicts an incorrect solution. A high number of local minima 
have appeared so that even the 
re-sampling method fails in this case to improve the results of the modelling. The local minima in the 1- or 
3-$\sigma$ 
error box on classical parameters provide as well wrong estimations of the 
stellar parameters (see columns 5 to 8 of Table~\ref{tablechi2-2}). In the bottom 
panel of Fig.~\ref{Fig3-Tefflogg-4freq}, the local minimum in the 1-$\sigma$ error box is located at the very 
limit of the box, 
and also leads to wrong estimations of the mass (16.6~$\msun$) and radius (9.43~$\rsun$).

Despite the poor results, the values of $\chi^2$ remain low, $\sim$0.02-0.03. The matching of the 
input modes by those of the grid is actually made easier. When a mode has its $\ell$ identified, it can only 
be matched to mode from the grid with a same $\ell$. Yet, when $\ell$ is unknown, this mode can then be fitted by 
modes from the grid with $\ell$=0 to 3, increasing the probabilities to match it with a mode of closer frequency, 
despite being of incorrect degree.

\begin{figure}
\begin{center}
\includegraphics[width=0.45\textwidth]{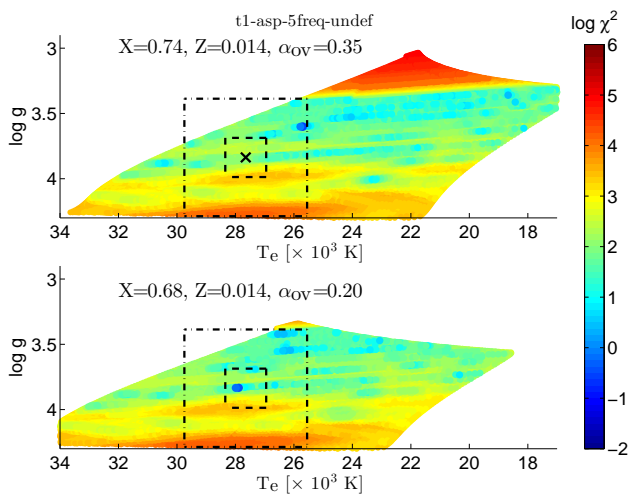}
\end{center}
\caption{Same as in Fig.~\ref{Fig1-Tefflogg-3freq}, now illustrated for the t1-asp-5freq-undef exercise.}
\label{Fig3-Tefflogg-5freq}
\end{figure}

With the addition of one frequency in the t1-$\ell$-5unknow test, the global minimum and the results of the re-sampling 
method still lead to incorrect 
inferences when only based on the seismic dataset (see columns 3 and 4 of Table~\ref{tablechi2-2}). In 
Fig.~\ref{Fig3-Tefflogg-5freq}, the number of local minima is now reduced, yet revealing the impact of a supplementary 
frequency.    
   
\bfseries Including the classical constraints. \mdseries With the help of the 1-$\sigma$ constraints, the solution is 
significantly improved; the input stellar parameters are now determined with good accuracy. Furthermore, the results 
from the re-sampling method (column 6 of 
Table~\ref{tablechi2-2}) 
present the same accuracy than in the t1-asp-5freq exercise. Yet, relaxing the constraint to the 3-$\sigma$ 
constraints, the predicted stellar parameters do no longer fit those of the t1 model.

A reasonably large set of frequencies, although unidentified, can however lead to an accurate asteroseismic modelling, 
provided it is  completed with tight constraints on the classical parameters. However, the stellar 
atmospheric parameters must be determined 
with a very good accuracy, as illustrated above when using the 
3-$\sigma$ error box on classical parameters. 

\subsection{Role of the input physics}
\label{section4-4}

We changed the physics of the model used as a target so that it differed from that of the grid. We 
examined the consequences on the quality of the global asteroseismic parameters, as well as on the constraints on the 
internal structure. We have seen in Sect.\ref{section4-1} that there is a significant degeneracy for the determination 
of overshooting and metallicity. This degeneracy will be particularly accentuated if the detailed chemical composition 
of 
the star differs from that used in theoretical models. For instance, we recalled in Sect.~\ref{section2-3} that between 
past determination of solar abundances like GN93 and the revision of AGS05, the metallicity decreased by $\sim$30\%. 
Hence we tailored the t2-gn93 exercise to test the impact of the adopted composition. We used for the target t2 
model the solar
mixture of GN93, while it is the AGS05 mixture that is adopted in the grid. The other properties of the t2 models are 
given in 
Table~\ref{table1-section4}. The set of input frequencies for the exercise are reported in Table~\ref{table2-section4} 
and is composed of six frequencies, with pairs of modes of degree $\ell_0$,$\ell_1$, and $\ell_2$.

\begin{table*}[!]
\caption{Results of the modelling for the t2-gn93 and t6-asp-diff exercises.}           
\label{tablechi2-3}     
\begin{center}  
\begin{tabular}{l l c c c c c c }        
\hline\hline                 
Exercise & Parameter & \multicolumn{2}{c}{No constraint} &  \multicolumn{2}{c}{1-$\sigma$ box constraint} &  
\multicolumn{2}{c}{3-$\sigma$ box constraint} \\    
 & & m.f. & MC & m.f. & MC & m.f. & MC \\
\hline        
  \smallskip
t2-gn93 & M (11) & 11.4 & 11.4$_{-0.1}^{+0.1\ddag}$ & 11.5 & 11.4$_{-0.3}^{+0.1}$ & 11.4 & 11.4$_{-0.1}^{+0.1\ddag}$ 
\\  
\smallskip
 & R (5.98) & 6.07 & 6.07$_{-0.03}^{+0.00\ddag}$ & 6.09 & 6.05$_{-0.04}^{+0.04}$ & 6.07 & 6.07$_{-0.03}^{+0.00\ddag}$\\
\smallskip 
 & X (0.70) & 0.70 & 0.70$_{-0.02\ddag}^{+0.02\ddag}$ & 0.74 & 0.72$_{-0.04}^{+0.02}$ & 0.70 & 
0.70$_{-0.02\ddag}^{+0.02\ddag}$ \\
\smallskip 
 & Z (0.016)  & 0.012 & 0.012$_{-0.002\ddag}^{+0.002}$ & 0.014 & 0.014$_{-0.002}^{+0.002}$ & 0.012 & 
0.012$_{-0.002\ddag}^{+0.002}$ \\
\smallskip 
 & $\alpha_{\mathrm{ov}}$ (0.20)  & 0.30 & 0.30$_{-0.05}^{+0.05}$ & 0.25 & 0.25$_{-0.10}^{+0.10}$ & 0.30 & 
0.30$_{-0.05}^{+0.05}$ \\
\smallskip 
 & $X_{\mathrm{c}} $ (0.351)  & 0.354 & 0.354$_{-0.001}^{+0.007}$ & 0.385 & 0.373$_{-0.043}^{+0.012}$ & 0.354 & 
0.354$_{-0.001}^{+0.007}$ \\
\smallskip 
 & $T_{\mathrm{e}}$ (25293)  & 26274 & 26274$_{-475}^{+0\ddag}$ & 25111 & -- & 26274 & -- \\
\smallskip 
 & $\log g$ (3.9258)  & 3.9288 & 3.9288$_{-0.0016}^{+0.0\ddag}$ & 3.9291 & -- & 3.9288 & -- \\
   \smallskip
 &  $\chi^2$ & 1.0338 & -- & 2.5574 & -- & 1.0338 & -- \\
  \smallskip
t6-asp-diff & M (10) & 10.2 & 10.2$_{-0.1\ddag}^{+0.1\ddag}$ & 10.2 & 10.2$_{-0.1\ddag}^{+0.1\ddag}$  & 10.2 & 
10.2$_{-0.1\ddag}^{+0.1\ddag}$  \\  
\smallskip
 & R (5.12) & 5.16 & 5.16$_{-0.00\ddag}^{+0.00\ddag}$ & 5.16 & 5.16$_{-0.00\ddag}^{+0.00\ddag}$ & 5.11 & 
5.16$_{-0.00\ddag}^{+0.06}$\\
\smallskip 
 & X (0.70) & 0.72 & 0.72$_{-0.02\ddag}^{+0.02}$ & 0.72 & 0.72$_{-0.02\ddag}^{+0.02\ddag}$ & 0.72 & 
0.72$_{-0.02\ddag}^{+0.02\ddag}$ \\
\smallskip 
 & Z (0.014)  & 0.016 & 0.016$_{-0.002\ddag}^{+0.002\ddag}$ &  0.016 & 0.016$_{-0.002\ddag}^{+0.002\ddag}$ & 0.016 & 
0.016$_{-0.002\ddag}^{+0.002\ddag}$ \\
\smallskip 
 & $\alpha_{\mathrm{ov}}$ (0\dag)  & 0.05 & 0.05$_{0.05\ddag}^{+0.05\ddag}$ & 0.05 & 0.05$_{0.05\ddag}^{+0.05}$ & 0.05 
& 
0.05$_{0.05\ddag}^{+0.05}$ \\
\smallskip 
 & $X_{\mathrm{c}} $ (0.388)  & 0.419 & 0.419$_{-0.000\ddag}^{+0.008}$ & 0.419 & 0.419$_{-0.000\ddag}^{+0.008}$ & 0.419 
& 0.419$_{-0.000\ddag}^{+0.008}$ \\
\smallskip 
 & $T_{\mathrm{e}}$ (24487)  & 24015 & 24015$_{-0\ddag}^{+365}$ & 24015 & -- & 24015 & -- \\
\smallskip 
 & $\log g$ (4.0196)  & 4.0212 & 4.0212$_{-0.0000\ddag}^{+0.0000\ddag}$ & 4.0212 & -- & 4.0212 & -- \\
   \smallskip
 &  $\chi^2$ & 2.5996 & -- & 2.5996 & -- & 2.5996 & -- \\ 
\hline
\end{tabular}
\end{center}
{Same comments as in Table~\ref{tablechi2-1}. Moreover, $\dag$ corresponds to input models with turbulent mixing: the 
extra mixing of the t6 models is equivalent to $\alpha_{\textrm{ov}}$~=~0.05.}
\end{table*}
 
\begin{figure}
\begin{center}
\includegraphics[width=0.46\textwidth]{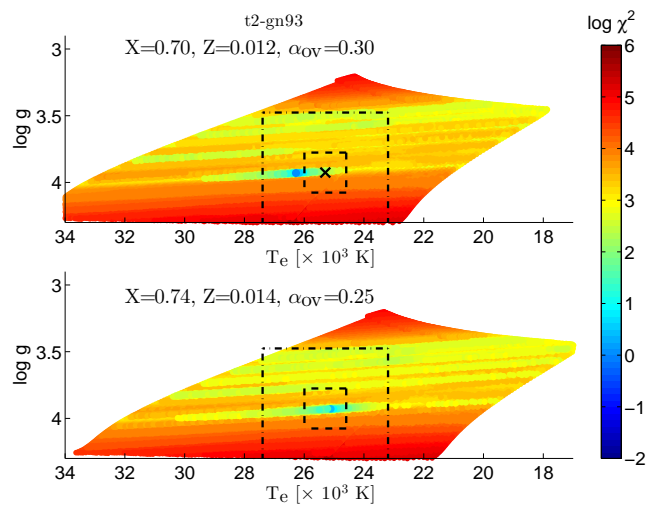}
\end{center}
\caption{Same as in Fig.~\ref{Fig1-Tefflogg-3freq}, but for the t2-gn93 exercise.}
\label{figchi2-77}
\end{figure}

The output parameters slightly overestimate the parameters of the t2 models with $M$=11.4~$\msun$ and 
$R$=6.07~$\rsun$. The best-fit model reproduces $\tau_{\textrm{dyn}}$ within 1\% the one of the t2 model, confirming 
the influence 
of radial modes on the modelling. We see in the top panel of Fig.~\ref{figchi2-77} that the lowest values of the merit 
function are indeed located in ridges, which actually correspond to places of equal $\tau_{\textrm{dyn}}$. The 
parameters $\alpha_{\textrm{ov}}$ and $Z$ are overestimated and underestimated, respectively (see column 3 of 
Table~\ref{tablechi2-3}). This comes as expected from the degeneracy in $\alpha_{\textrm{ov}}$ and $Z$, 
accentuated by the difference in the chemical mixture. This effect was already observed in the modelling of HD129929, 
then from an ad hoc change in the mixture made by the authors \citep{thoul04}. Actually, in reason of the predominance 
of nuclear energy 
production by CNO cycle in B stars, for a given X and Z, a GN93 model will be more luminous than an AGS05 one. The 
reason is that C, N, and O are more abundant constituents of the metal mixture in the first case. Since we try to model 
a GN93 model with 
AGS05 models, the Z and 
$\alpha_{\textrm{ov}}$ likely adapt and do not correspond to those of the GN93 model. 

\bfseries Including the classical constraints. \mdseries With predicted $Z$~=~0.014 and $\alpha_{\textrm{ov}}$~=~0.25 
(using 1-$\sigma$ error box, see column 5 of Table~\ref{tablechi2-3}), the discrepancy between the solution and the t2 
model is reduced, but $X$ is now clearly overestimated, illustrating again the 
degeneracy on the overshooting and chemical composition determinations. The results of the re-sampling method are not 
very 
different (column 6 of Table~\ref{tablechi2-3}), although the errors on the parameters are larger and so
enclose the actual values of the t2 model. In Fig.~\ref{figchi2-77}, the 
global minimum and the lowest values of $\chi^2$ surrounding it, lie outside the 1-$\sigma$ error box on the classical 
parameters (top panel). In the bottom 
panel, the local minimum found in the 1-$\sigma$ error box is not well defined and its surrounding $\chi^2$ 
values are very similar: the solution is not well constrained, resulting in larger uncertainties on the derived 
parameters. The 3-$\sigma$ constraints do not bring any information as the global 
minimum lies within it. This global minimum is clearly 
brighter and hotter than the target t2 model.

         \begin{figure}
\centering
\includegraphics[width=0.48\textwidth]{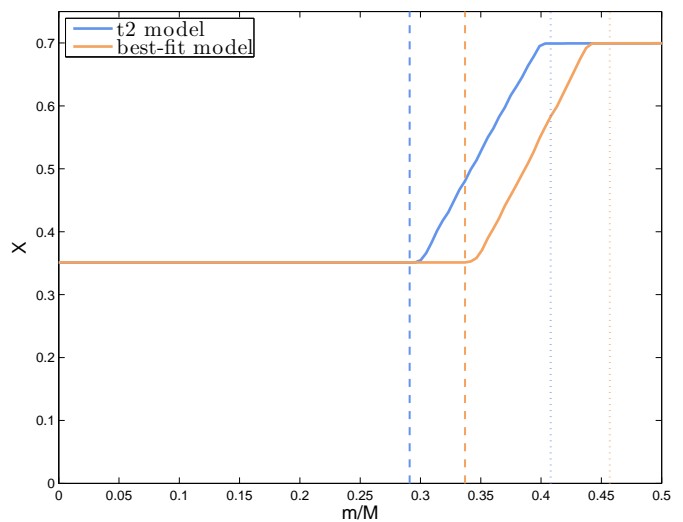}
      \caption{Same as in Fig.~\ref{Fig1-mm-5freq}, but for the t2 model and its best-fit model from the 
t2-gn93 exercise.}
         \label{Fig1-mm-t2}
   \end{figure}

Overall, the inaccuracy induced by the difference in the chemical mixture remains limited to a few \% on global 
parameters such as $M$ and 
$R$. However, it is difficult to estimate the accuracy of parameters such as metallicity and 
overshooting given the degeneracy and compensatory role they play. Verifying how the limits of the 
convective core and $\nabla \mu$ are reproduced would better hint at the robustness of the asteroseismic modelling on 
this question. We present in Fig.~\ref{Fig1-mm-t2} the internal profile of $X$ of the t2 and its seismic best-fit 
models, showing that the size in mass of the fully mixed regions and $\nabla \mu$ are not retrieved by the modelling 
(accuracy errors of 16 and 12\%, respectively). Moreover, one mode ($k_1$, $\ell_1$) of the input dataset is mixed with 
a dominant g-character, in principle an optimal information on these stellar layers.  Yet, we see from 
Table~\ref{tablechi2-11} that the accuracy on these limits is much better in terms of $r/R$, $\sim$5\%. It 
highlights that the asteroseismic information is defined in terms of variables related to oscillations, so essentially 
sensitive to $r$. While the acoustic structure of the star can be well reproduced, this is not necessarily the case of 
the mass distribution in the layers. Here it is clearly the difference in the chemical mixture that hinders the 
recovery of mass of the convective core and its overshooting region.

We did two other tests where the target models were computed with the GN93 mixture and with OPAL opacities (versus OP 
in 
the grid), but in one case, no overshooting was included. In both cases, the additional change in the physics did not 
alter precedent conclusion: we were still able to infer $M$ and $R$ with a good accuracy. But in the case with 
overshooting, we faced the same degeneracy on the chemical composition and overshooting determination. It resulted in 
an important discrepancy of $\sim$20\% on the $m/M$ limit of the fully-mixed central region. Nevertheless, in 
the case without overshooting, the solution correctly retrieved the absence of overshooting (but failed at the 
chemical composition). The limits on the convective core were still of $\sim$5\% in terms of $r/R$, but remained of 
the same order when expressed in $m/M$.
   
\subsection{Nature of extra mixing}

We have considered insofar models where the extra mixing was treated with a classical instantaneous 
prescription \citep[see e.g.][]{maeder75}. Other processes can be responsible for extra mixing near the convective 
core, as for example turbulent mixing induced by rotation. Although this latter process has almost the same impact on 
the 
stellar evolutionary tracks as the overshooting \citep[see e.g.][]{talon97}, the process acts like a diffusive process 
leading to smoother chemical composition gradients at the boundary of the convective core. The effect is in 
principle noticeable in comparison to the sharp profile generated in that same region by a very efficient mixing.

We explored this question in the t6-asp-diff exercise, in which turbulent mixing is implemented as 
a diffusive process parametrised by a turbulent diffusion coefficient, $D_{\textrm{t}}$. We computed the t6 target 
model (see Table~\ref{table1-section4}) with a value of $D_{\textrm{t}}$~=~70,000~cm.s$^{-2}$, calibrated to correspond 
to a model computed with the Geneva code \citep{genevacode} with an initial equatorial rotational velocity on the ZAMS
($V_{\textrm{i}}$) of 50~km/s. Doing so, the t6 model is equivalent (in terms of the extent of the central fully-mixed 
region) to a model without diffusive mixing but 
$\alpha_{\textrm{ov}}$~=~0.05.

The input frequency set is composed of two radial modes, three $\ell_1$, and two $\ell_2$ modes. The modelling with 
these 
seismic constraints recovers with good accuracy $M$, $R$, and $\alpha_{\textrm{ov}}$ (see column 
3 of Table~\ref{tablechi2-3}), but $X$ and $Z$ are overestimated.

\begin{figure}
\begin{center}
\includegraphics[width=0.45\textwidth]{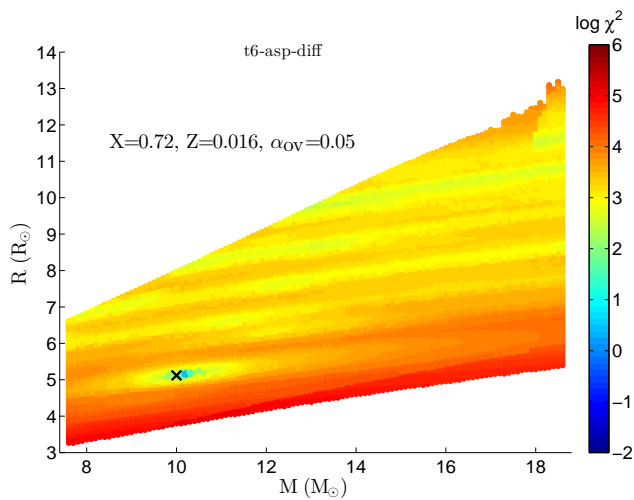}
\end{center}
\caption{Same as in Fig.~\ref{Fig1-MR-3freq}, but now only illustrated for the parameters ($X$, $Z$, 
$\alpha_{\textrm{ov}}$) of the global minimum of the t6-asp-diff exercise.}
\label{figchi2-97}
\end{figure}

The re-sampling method does not predict different results and it fails to determine consistent errors on 
the parameters (see column 4 of Table~\ref{tablechi2-3}). The reason is that the global minimum is deeply marked in the 
$\chi^2$ map in Fig.~\ref{figchi2-97}, so that other solutions hardly emerge from the MC simulations. The global 
minimum lies within the 1-$\sigma$ error box on $T_{\textrm{e}}$--$\log g$, and would do the same 
considering typical 1-$\sigma$ observational error bars on $Z/X$. The classical constraints appear as no help in this 
case. 

\begin{figure}
\centering
\includegraphics[width=0.48\textwidth]{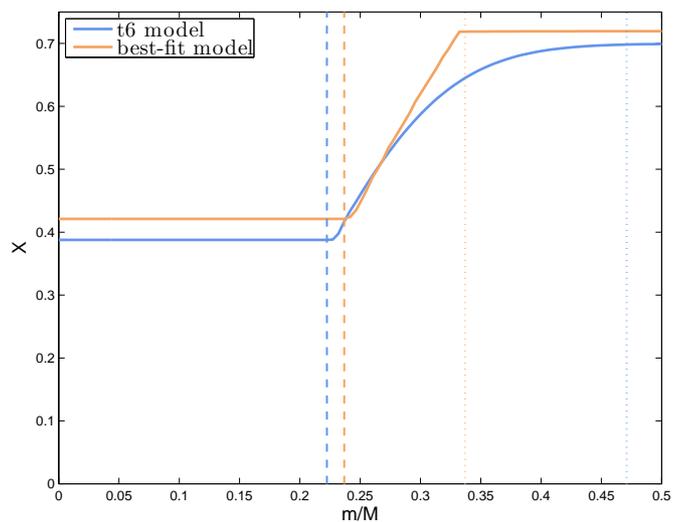}
      \caption{Same as in Fig.~\ref{Fig1-mm-5freq} but for the t6 model and its best-fit model from the 
t6-asp-diff exercise.}
         \label{Fig1-mm-t5}
   \end{figure}

Of prime importance, we finally look at the limits of the convective core and $\nabla \mu$ in Fig.~\ref{Fig1-mm-t5}. 
The limit of the 
convective core is determined with a precision of ~5\% (see also Table~\ref{tablechi2-11}) in $m/M$, as it could be 
expected, since the t6 target model and the grid share the same chemical mixture (AGS05). However, the 
modelling fails at determining the location of $\nabla \mu$, and so to be sensitive to the nature of the extra mixing. 
This is somehow expected as the grid did not include any diffusive mixing, confirming that getting constraints on the 
internal processes depends on the stellar models used. It suggests that a diffusive mixing 
should be as well considered in the 
grid for assessing its presence and efficiency in $\beta$ Cephei stars. This would add a degree of freedom, under the 
form 
of a diffusive coefficient taken as an additional parameter to adjust. \citet{lovekin10} tested the addition of a 
parameter in the fit, although applied to retrieve the rotation velocity of the star. Their results were encouraging 
as the convergence towards a reliable modelling of $\beta$ Cephei stars was not hindered by the additional parameter to 
fit.

We also did another test (not detailed here) considering a target model with diffusive mixing and the GN93 mixture, 
still with the grid only including instantaneous mixing. As expected, the inaccuracy increased, in particular on 
the determination of the convective core. Its inferred mass was then underestimated by 15\%. The solution also exceeded 
by 15\% the limit in $m/M$ on $\nabla \mu$, confirming the insensitivity to the nature of extra mixing with an 
inappropriate model grid. Despite the internal structure was less-well characterised, the global stellar parameters 
were 
recovered with the same accuracy and precision than it the tests where the target stars included no diffusive mixing 
and 
the GN93 mixture in Sect.~\ref{section4-4}.
   
\section{Conclusion}

The asteroseismology of $\beta$ Cephei stars has already been applied in several occasions on well-characterised 
targets \citep[see review by][]{aerts15}, leading to the derivation of the global parameters of these massive stars. 
These attempts also confirmed the potential for constraining internal processes such as rotation or mixing at the 
convective core boundary. With the latest results of the BRITE space mission, new seismic data extend the 
constraints for some well-known beta Cephei stars \citep{dd17,handler17,walzak19}. The first results of this mission 
have focused on 
the excitation of pulsation modes, confirming the call for a revision of stellar opacities, as previously stated 
from earlier studies on $\beta$~Cephei stars with fewer modes 
\citep{dziembo08,montalban08a,salmon12,walzak13,dd13,cugier14}. The new 
data also give the 
opportunity to improve the snapshot of the internal structure that can be deduced with asteroseismology.

In this work, we explored in detail the role of the quality of the seismic data on precision and accuracy in the
derivation of the stellar parameters and internal structure. We have developed a method that can be systematically 
applied to the modelling of $\beta$ Cephei stars. It is based on a 
re-sampling of the observed seismic frequencies, using Monte-Carlo simulations. It allows for instance to 
methodically estimate the errors on the derived parameters.

We applied this method in a series of hare and hound exercises, for which we simulated $\beta$ Cephei targets using 
theoretical stellar models. These tests aimed at defining the conditions required to obtain reliable and accurate 
seismic solutions. We also carefully characterised the dependence 
on the physics of stellar models used for the modelling. We explored the potential to determine the limits of the 
central mixed regions (convective core and overshoot region) as 
well as the nature of additional mixing processes at the convective core boundary. This would complement recent 
results obtained for SPBs \citep[e.g.][]{may21}, stars with 
a similar structure albeit less massive. The analysis of the exercises shows that:
\begin{itemize}
 \item Ideally the set of frequencies used for the modelling of a $\beta$~Cephei star should include at least four to 
five 
frequencies, 
with the knowledge of their angular degree ($\ell$). Depending on the presence of mixed modes and the addition of 
fundamental parameters from non-seismic observables, accurate asteroseismic modelling with fewer modes is still 
possible, but is very 
dependent on the modes detected. The misidentification of one $\ell$ degree hindered to retrieve correctly the stellar 
parameters: but the value of the merit function was then significantly degraded, signalling the issue. 
\item In the absence of identification of the modes, a set of four or five frequencies is not sufficient to 
determine the stellar parameters on the sole basis of the seismic dataset. Provided some
fundamental parameters (effective temperature, surface gravity) are known from non-seismic constraints, we are able to 
retrieve the original stellar properties from a set of five frequencies.
 \item If the nature of the extra mixing is the same between the star and the theoretical grid, the limit 
of the chemically homogeneous central regions (convective core plus overshoot region) is 
inferred with a good accuracy in terms of acoustic variables, like the radius. The extent of this region is
correctly retrieved in terms of the mass provided the chemical mixture of the models is representative of the 
star. Therefore, the knowledge 
not only of the metallicity, but also of the individual chemical element abundances is required to identify the 
chemical mixture to be adopted. This 
encourages to systematically carry observations for determining the detailed abundances of $\beta$ 
Cephei stars.
\item When the nature of the extra mixing differs between the star and the theoretical grid, determining the size in 
mass of the convective core remains accurate when the chemical mixture of the grid reproduces that of the star. 
However, determining the limit of the chemical composition gradient and so the nature of the extra mixing appears 
unsuccessful. It hints at including in the analysis theoretical models with 
different treatments of the extra-mixing processes. Refined seismic diagnosis tools must be developed in that 
specific aim.
\end{itemize}

The re-sampling method based on Monte-Carlo method has demonstrated its capability to improve the results of the 
modelling when the solutions were initially poorly constrained. It also appears reliable to deliver realistic 
errors on the inferred parameters, as in most cases its error intervals were predicting correctly the real parameters 
of the target stellar models. It would be interesting to carry out further comparison with other statistical 
indicators in the future. For instance, we could perform a parallel analysis with the Mahalanobis distance, which use 
was proposed for asteroseismology of SPB pulsators by \citet{aerts18}. This latter accounts directly for 
correlation between the parameters by including the covariances in its expression. The errors it predicts would be 
compared to those of our method. Yet, to compute correctly the covariance matrix, the required density and size of the 
grid of models have first to be estimated. As our method enables us to refine the region of the 
parameter space enclosing the solution, we could also use it as an exploratory method for providing initial guesses of 
a local method, such as the Levenberg-Marquardt algorithm (a two-step approach which was already followed for 
red-giants, e.g. \citealt{buldgen19} or solar-like pulsators, e.g. \citealt{farnir20}).

\begin{acknowledgements}
S.J.A.J.S., P.E., F.M. and G.M. have received funding from the 
European Research
Council (ERC) under the European Union’s Horizon 2020 research and innovation programme (grant agreement No 833925, 
project STAREX). J.M. and A.M. acknowledge support from the European Research Council Consolidator Grant funding scheme 
(project ASTEROCHRONOMETRY, G.A. n. 772293, http://www.asterochronometry.eu).
G.B. acknowledges funding from the SNF AMBIZIONE
grant No. 185805 (Seismic inversions and modelling of transport processes in
stars).
\end{acknowledgements}

\bibliographystyle{aa}
\bibliography{biblio}

\end{document}